\documentclass{usiinftr}

\usepackage{amsmath} 
\usepackage{amsfonts} 
\usepackage{bold-extra} 
\usepackage[T1]{fontenc} 
\usepackage[scaled]{beramono}

\let\emptyset\varnothing

\usepackage{listings}
\usepackage[ruled,vlined]{algorithm2e}
\DontPrintSemicolon
\renewcommand{\CommentSty}[1]{\textnormal{#1}}
\SetKwComment{tcp}{$\triangleright$ }{}
\SetVlineSkip{0cm}
\SetAlgoSkip{}

\usepackage{alltt}
\usepackage{color}
\usepackage{url}
\usepackage{graphicx}
\usepackage{multirow}
\usepackage{blkarray}
\usepackage{balance}
\usepackage{mdframed}

\makeatletter \def\url@leostyle{\@ifundefined{selectfont}{\def\UrlFont{\sf}}{\def\UrlFont{\scriptsize\ttfamily}}} \makeatother
\begin{document}

\title{Optimized Disk Layouts for\\ Adaptive Storage of Interaction Graphs}

\author{Robert Soul\'{e}}{1}
\author{Bu\u{g}ra Gedik}{2}

\affiliation{1}{\USIINF}
\affiliation{2}{Department of Computer Engineering, Bilkent University, Turkey}

\TRnumber{2014-04}

\maketitle
\begin{abstract}
  We are living in an ever more connected world, where data recording the
interactions between people, software systems, and the physical world is
becoming increasingly prevalent. This data often takes the form of a
temporally evolving graph, where entities are the vertices and the
interactions between them are the edges. We call such graphs interaction
graphs. Various application domains, including telecommunications, transportation, and social
media, depend on analytics performed on interaction graphs. The ability to
efficiently support historical analysis over interaction graphs require
effective solutions for the problem of data layout on disk. This paper
presents an adaptive disk layout called the railway layout for optimizing disk
block storage for interaction graphs. The key idea is to divide blocks into
one or more sub-blocks, where each sub-block contains a subset of the
attributes, but the entire graph structure is replicated within each
sub-block. This improves query I/O, at the cost of increased storage overhead.
We introduce optimal ILP formulations for partitioning disk blocks into
sub-blocks with overlapping and non-overlapping attributes. Additionally, we
present greedy heuristic approaches that can scale better compared to the ILP
alternatives, yet achieve close to optimal query I/O. To demonstrate the
benefits of the railway layout, we provide an extensive experimental study
comparing our approach to a few baseline alternatives.
            
\end{abstract}

\section{Introduction}\label{sec:introduction}
\noindent
We are living in an ever more connected world, where the data generated by
people, software systems, and the physical world is more accessible than
before and is much larger in volume, variety, and velocity. In many
application domains, such as telecommunications, transportation, and social media, live data
recording the interactions between people, systems, and the environment is
available for analysis. This data often takes the form of a temporally
evolving graph, where entities are the vertices and the interactions between
them are the edges. We call such graphs \emph{interaction graphs}. 

Data analytics performed on interaction graphs can bring new business insights
and improve decision making. For instance, the graph structure may represent
the interactions in a social network, where finding communities in the graph
can facilitate targeted advertising. In the telecommunications (telco) domain,
call details records (CDRs) can be used to capture the call interactions
between people, and locating closely connected groups of people can be used
for generating promotions. 

Interaction graphs are temporal in nature, and more importantly, they are
append-only. This is in contrast to relationship graphs, which are updated via
insertion and deletion operations. An example of a relationship graph is a
social network capturing the follower-followee relationship among users.
Examples of interactions graphs include CDR graphs capturing calls between telco
customers or mention graphs capturing interactions between users of a
micro-blogging service, like Twitter. The append-only nature of the interaction
graphs make storing them on disk a necessity. Furthermore, the analysis of this
historical interaction data forms an important part of the analytical landscape.

 Since interaction graphs can potentially grow forever, they present a storage
challenge for system designers.  Even on modern servers with large amounts of
memory, system designers cannot assume that the entire graph can fit. Instead,
interaction graph systems must store their data on disk.

The ability to efficiently support historical analysis over interaction graphs
requires effective solutions for the problem of \emph{data layout} on disk.
Most graph algorithms are characterized by locality of
access~\cite{hoque12}, which is a direct result of the traversal-based
nature of most of the graph algorithms. This is often taken advantage of by
co-locating edges in close proximity within the same disk
blocks~\cite{g-store}. This way, once a disk block is loaded into main
memory buffers, several edges from it can be used for processing, reducing the
disk I/O. 

In interaction graphs, the locality of access is even more pronounced. First, 
the analysis to be performed on the interactions can be restricted to a
temporal view of the graph, such as finding the influential users over a given
week of interactions. This means that edges that are temporally close are accessed together. Second, traversals are again key to many graph
algorithms, such as connected components, clustering coefficient, PageRank,
etc. This means that edges that are close by in terms of the path between
their incident vertices as well as their timestamps should be located together
with the same blocks.  In our earlier work~\cite{gedik14}, we introduced an
interaction graph database that works on this principle of access locality. It
uses a disk organization that consists of a set of blocks, each containing a
list of \emph{temporal neighbor lists}. A temporal neighbor list contains a
head vertex and a set of incident edges within a time range. The layout
optimizer aims at bringing together, into the same disk block, temporal
neighbor lists that are ($i$) close in terms of their temporal ranges, ($ii$)
have many edges between them, and ($iii$) have few edges going into temporal
neighbor lists outside the block.

Many real world graph databases contain attributes. In the case of interaction
graphs, the attributes can be considered as properties associated with the
edges representing the interactions. Attributes can be stored in two ways,
either separately (e.g., in a relational table), or locally with the temporal
neighbor lists.  If they are stored separately, then the graph database cannot
take full advantage of locality optimizations performed for block
organization. The database must go back and forth between the disk blocks to
access the edge attributes.  On the other hand, if attributes are stored
locally in the disk blocks containing the graph structure, then there can be
significant overhead due to disk I/O if only a few attributes are needed to
answer a query. 

To query an interaction graph, most algorithms traverse the graph structure to
access the relevant attributes.  Frequently, there are correlations among the
attributes accessed by different queries. For example, queries $q_1$ and $q_5$
might access attributes $a_1$ and $a_2$, while queries $q_2$, $q_3$ and $q_4$
access attributes $a_3$ and $a_4$. Because interaction graphs are temporal,
the co-access correlations for the attributes can vary for different temporal
regions.  Moreover, the co-access correlations might be unknown at the
insertion time, but be discovered later, when the workload is known.

It is widely recognized that query workload and disk layout have a significant
impact on database performance~\cite{alagiannis14,grund10,stonebraker05}.  For
table-based relational databases, this fact has led database designers to
develop alternative approaches for storage layout: row-oriented
storage~\cite{rowOrg} is more efficient when queries access many attributes
from a small number of records, and column-oriented storage~\cite{colOrg} is
more efficient when queries access a small number of attributes from many
records~\cite{stonebraker05}.  Unfortunately, although interaction graph
databases, like relational databases, are the target of diverse query
workloads, there is no clear correspondence to a row-oriented or
column-oriented storage layout.

This paper presents an adaptive disk layout called the \emph{railway layout}
and associated algorithms for optimizing disk block storage for interaction
graphs. The key idea is to divide blocks into one or more sub-blocks, where
each sub-block contains a subset of the attributes (potentially overlapping),
but the entire graph structure is replicated within each sub-block. This way,
a query can be answered completely by only reading the sub-blocks that contain
the attributes of interest, reducing the overall I/O. 

There a number of challenges in achieving an effective adaptive layout. First,
we need to find the partitioning of attributes that minimizes the query I/O.
To address this, we model the problems of overlapping and non-overlapping
attribute partitioning as mixed-integer linear programs (ILPs), and provide
optimal solutions that minimize the query I/O cost. Second, the query
workload, and thus the attribute access pattern can change over time. For this
purpose, our railway layout supports customization of the attribute
partitioning of sub-blocks on a per-block basis. Third, such flexibility
necessitates online configuration of attribute partitioning as the query
workload evolves, which in turn requires fast algorithms for performing the
attribute partitioning. For this purpose, we develop greedy heuristic
algorithms for both overlapping and non-overlapping partitioning scenarios.
These algorithms can scale to larger number of attributes, yet provide close
to optimal query I/O performance. Finally, the railway layout trades off
storage space to gain improved query I/O performance. The storage overhead is
more pronounced for the case of overlapping partitioning. To address this, we
limit the amount of storage  overhead that can be tolerated, and integrate
this limit to both our ILP formulations, as well as our greedy heuristics.

Our experiments demonstrate the benefits of the railway layout.  For a storage
increase of just 25\%, the optimal overlapping partitioning algorithm reduces
the query I/O cost by 45\%.  When allowed to double the storage usage, the
overlapping partitioning algorithm can reduce the I/O cost by 73\%. The
heuristic algorithm performs almost as well, reducing the I/O cost by 72\%, but
reduces the running time needed to find a solution by orders of magnitude.

In summary, this paper makes the following contributions:
\begin{itemize}
\item We introduce the railway layout for adaptive organization of interaction
graphs on disk. 
\item We introduce optimal ILP formulations for partitioning disk blocks into 
sub-blocks with overlapping and non-overlapping attributes, given a query
workload. Our formulation also support upper bounding the amount of storage
overhead introduced as a result of the railway layout.
\item To support online adaptation, we develop greedy heuristics that can scale
better compared to the ILP alternatives, yet achieve close to optimal query
I/O.
\item We provide an extensive experimental study comparing our approach to a
few baseline alternatives.
\end{itemize}

The rest of the paper is organized as follows. Section~\ref{sec:system} gives
an overview of the railway layout in the context of an interaction database
and motivates its design. Section~\ref{sec:optimal} formalizes the optimal
railway layout design problem. Section~\ref{sec:ilp} gives Mixed Integer
Linear Programming formulation of the optimal layout for overlapping and
non-overlapping scenarios. Section~\ref{sec:heuristic} introduces our
heuristic solutions for the same. Section~\ref{sec:evaluation} presents an
experimental evaluation of our system. Section~\ref{sec:related} discusses
related work and Section~\ref{sec:conclusion} concludes the paper.



\begin{figure}[!Ht]
\centerline{\includegraphics[width=0.6\columnwidth]{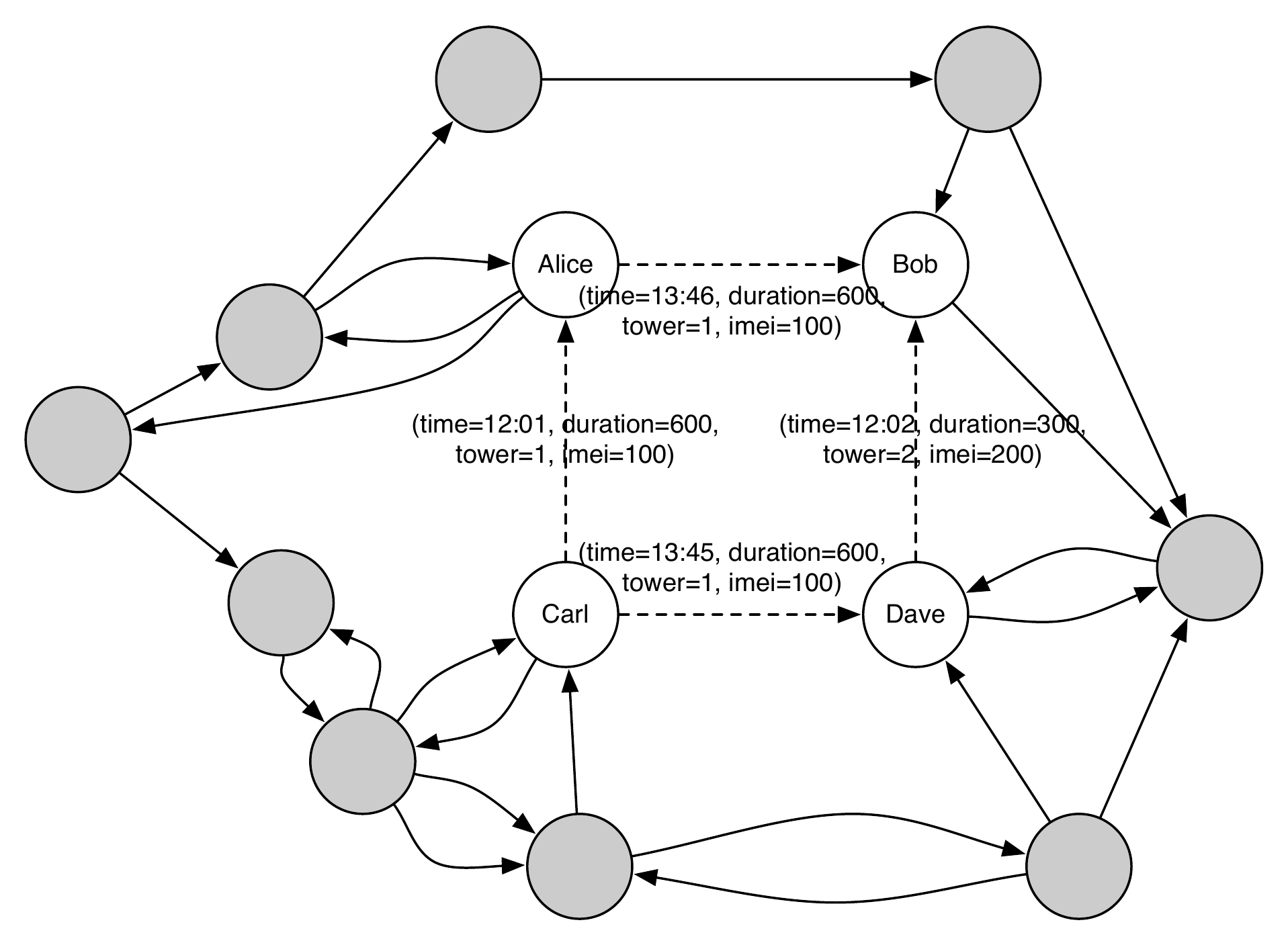}}
 \caption{A partial example interaction graph for call data records, capturing the
   telephone calls among a set of people. To motivate our design,
   this paper will focus on a subgraph at a particular time range, indicated
   by the nodes colored white. Each edge in the graph is
   associated with attributes for the interaction, including the time the call
   was placed, the duration of the call, the cell phone tower, and the IMEI
   number identifying the device used.}
 \label{fig:example}
 \end{figure}

\section{System Overview}\label{sec:system}
\noindent
The design of the railway disk layout builds on our prior work~\cite{gedik14}, 
which organized the disk layout for interaction graph databases to improve
access locality. The railway layout extends the earlier design, by enabling
the system to adapt the layout to changing  workloads, with the goal of
reducing the disk I/O during querying, in exchange for a slight increase in the
disk space used to store the graph.

\subsection{Motivating Example}
\noindent
To explain the design of the railway layout, we first introduce a small,
motivating example. Figure~\ref{fig:example} shows a graph for the telephone
call interactions among a set of people. Each node in the graph represents a
person, and each edge in the graph represents a phone call from a caller to
a callee. Each edge is associated with a set of attributes that maintain the
details of the interaction, including the time the call was placed, the duration
of the call, the cell phone tower, and the IMEI number identifying the device
used to place the call. Thus, the schema for each edge is as follows:
\begin{alltt}
\centering call(time, duration, tower, imei)
\end{alltt}

Recall that interaction graphs are append-only, and evolve over time. In other
words, new timestamped edges are continuously added to the graph. For
explanatory purposes, we focus on a subset of a graph at a particular time
range. In the figure, the subset is indicated by the white nodes, and the edges
between them. In this subset, there were four call interactions. One of them
was a call from Alice to Bob, starting at 13:46. They spoke for 600 seconds.
The call was received by cell phone tower 1, and the Alice's phone had an IMEI
number of 100. 

A telecommunications company performs various analytics by processing
the graph. For example, in order to understand how they should price their service
plans, they might want to capture the duration of all calls for each user. To
plan for infrastructure provisioning, they might want to record a count of the
number of calls that each cell phone tower handled.

In an interaction graph, queries are associated with a time range, $[t_{start},
t_{end}]$.  To answer queries, the graph database system must traverse the
subgraph that contains edges with timestamp $t$, such that $t_{start} \leq t
\leq t_{end}$. As the system traverses the graph, it reads the relevant
attributes to answer the query. Note that a query might access all or some of the
attributes. As concrete examples, imagine we have two queries. Query \texttt{q1}
asks for the average duration for calls from each tower. Query \texttt{q2} asks
for the count of calls made by each type of device. In other words we say that
each query accesses a subset of the attributes:%
\begin{alltt}
\centering q1 = \{duration, tower\}, q2 = \{imei\}
\end{alltt}

\begin{figure}[Ht]
\centering
\includegraphics[width=0.6\linewidth]{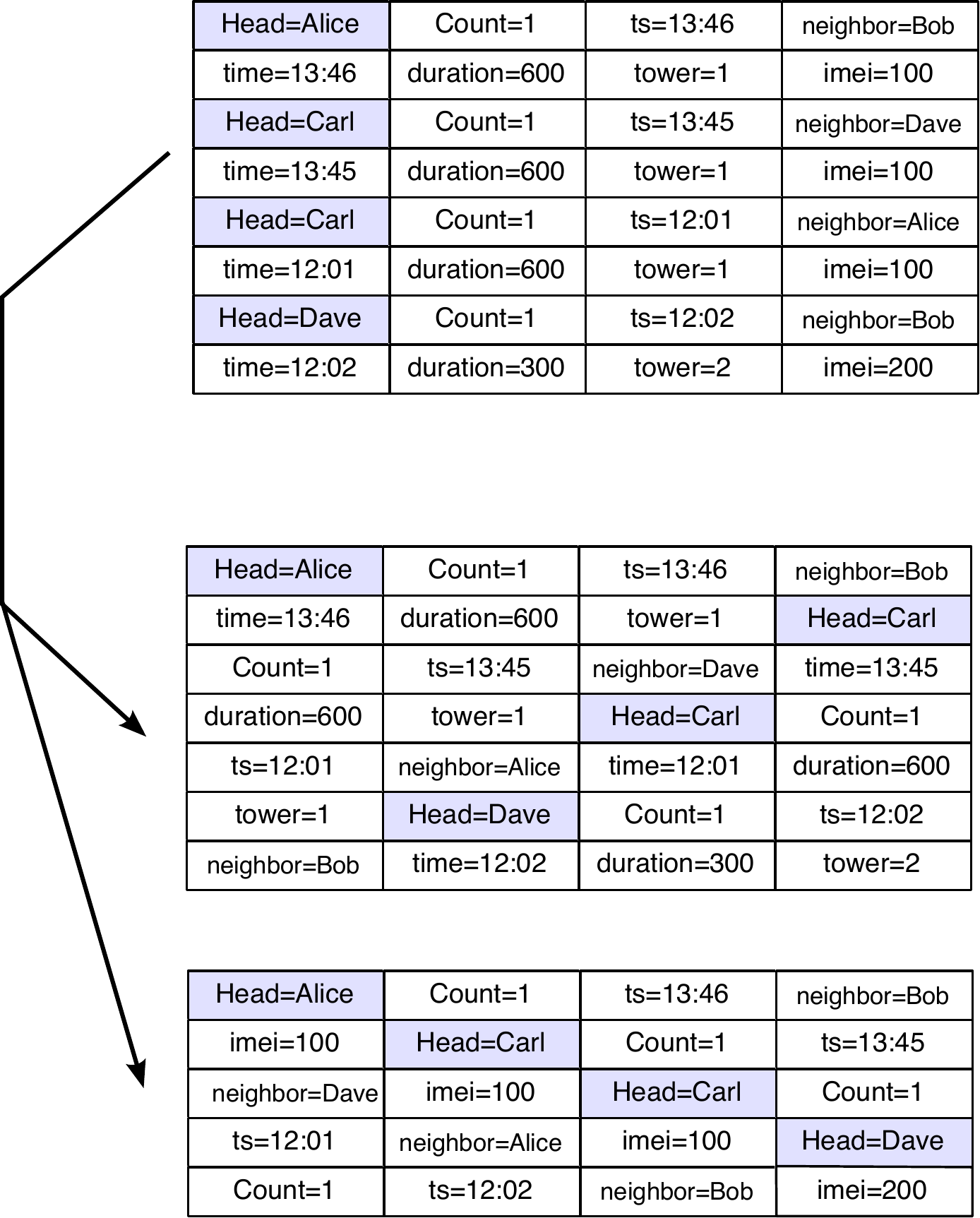} 
 \caption{The standard disk block storage for an interaction graph, and a
   partitioning into sub-blocks for the railway layout. Each sub-block maintains
 its own copy of the neighbor list, and a subset of the attributes.}
 \label{fig:before_after}
 \end{figure}

\subsection{Storage for Locality}
\noindent
There are several ways in which one might store a graph on disk. The graph
structure can be stored as a matrix representation or an adjacency list. Most
graph databases choose an adjacency list representation because they reduce
the storage overhead when graphs are sparse, and it is faster to iterate over
the edges when traversing the graph.  Attributes associated with each edge
could be stored separately in a relational table, or along with the edges.
Storing the attributes with the edges improves the locality, since the
database can read the graph structure and associated attributes from the same
disk block. To improve locality, typical disk layout schemes try to group as
many adjacent nodes as possible in the same disk block.

Building on this basic design, our prior work~\cite{gedik14} extended the
notion of locality to include a temporal dimension for handling interaction
graphs. Nodes are placed in the same block if they are close together both
spatially and temporally. Based on the edge timestamps, the adjacency lists
are divided into multiple pieces and based on closeness of the nodes within
the graph, these partial adjacency lists are combined into blocks. The
locality of a block is determined by its \emph{conductance} (i.e., the ratio
of the number of dangling half edges), and its \emph{cohesiveness} (i.e., a
metric used to find highly connected components). Our earlier work describes a
greedy algorithm for forming disk blocks with respect to this notion of
locality.  

Once the algorithm divides the graph into disk blocks, the graph data and
attributes are stored in the layout scheme illustrated in the top of
Figure~\ref{fig:before_after}.  Note that this is an adjacency list
representation in which attributes are stored with the edges.  Each disk block
contains a sequence of vertices, identified by a \emph{head-node id}, followed
by a \emph{count} of the number of neighbors, and then the neighbor list
itself. Each entry in the neighbor list is composed of a \emph{timestamp}, an
\emph{id} for the destination vertex, and the properties for that edge.  In
the top of Figure~\ref{fig:before_after}, all of the information
from the example interaction subgraph is stored in a single disk block. 

\begin{figure*}[ht]
\centering
\includegraphics[width=0.7\textwidth]{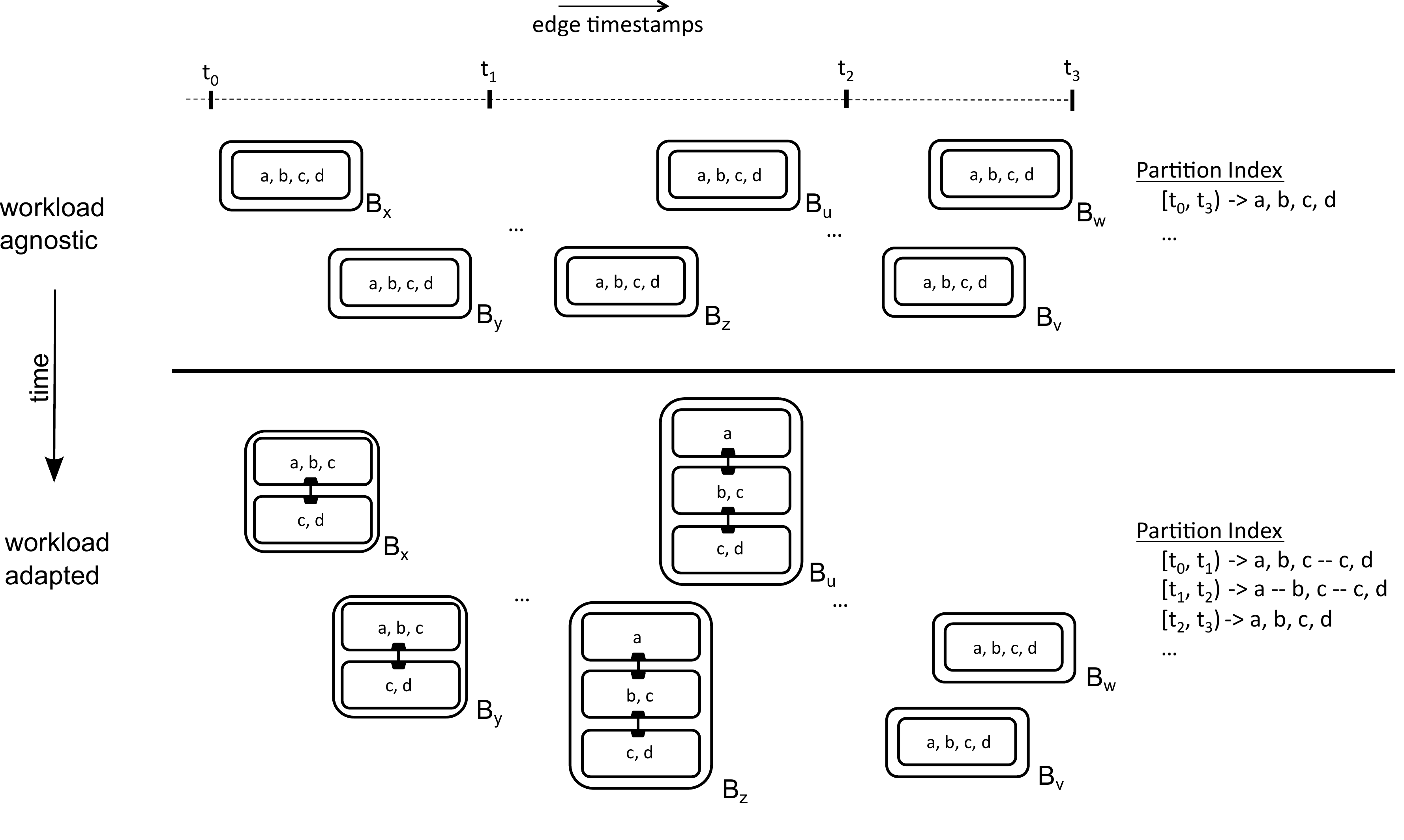} 
 \caption{A database system implementing the railway layout will adapt the disk
   storage over time.}
 \label{fig:rail_layout}
 \end{figure*}

\subsection{Railway Layout}
\noindent
This paper introduces a new disk layout scheme, called the \emph{railway
layout} illustrated in the bottom of Figure~\ref{fig:before_after}.
With the railway layout scheme, blocks are partitioned into sub-blocks, such
that each sub-block contains the adjacency list representation from the
original block, but only a subset of the attributes. The subset of attributes
assigned to each sub-block is determined by the query workload.

For example, given queries \texttt{q1} and \texttt{q2}, the railway layout
would store the attributes \texttt{time}, \texttt{duration}, and
\texttt{tower} in one sub-block, and the attribute \texttt{imei} in a second
sub-block. In the ideal case, a query can be answered completely by reading a
single sub-block that contains only the relevant information, and none of the
irrelevant information, reducing the overall I/O cost. Of course, this layout
comes at the expense of storage, as the graph structure information is
duplicated in each sub-block. We argue that in general, I/O cost is more
important than storage overhead, because a certain level of storage overhead
can be accommodated by adding additional disks. In Sections~\ref{sec:optimal}
and~\ref{sec:heuristic}, we will present our optimal and heuristic algorithms
for discovering the sub-block partitions that keep the overhead below a
user-specified threshold, while minimizing the disk I/O for the queries.

\subsection{Adaptation}
\noindent
Because interaction graphs are append-only, and new edges are continuously
added, there is a unique opportunity to adapt the disk layout with changing
workloads over time. A database system utilizing the railway layout design
would continually monitor the workload, and re-adjust the disk layout for
historical data. This is illustrated in Figure~\ref{fig:rail_layout}. In the
figure, we have an interaction graph with 4 attributes, namely $a$, $b$, $c$,
and $d$. Initially, without any workload optimization, all disk blocks have a
single sub-block  that contains the entire set of attributes. This is shown in
the upper half of the  figure. After some time, the database adapts to the
workload. This is shown in the lower half of the figure. We see that blocks
from different time ranges have adapted differently, as the workload they
observe is different. For  instance, blocks $B_X$ and $B_Y$ were partitioned
into two sub-blocks as $a, b, c - c, d$, whereas blocks $B_Z$ and $B_U$ were
partitioned into three sub-blocks as $a - b, c - c, d$,  and the blocks $B_V$
and $B_W$ stayed intact as $a, b, c, d$. A partition index is kept to track
the partitioning of blocks in different time regions of the interaction graph
database, which is shown on the right-hand side of the figure. 

In the rest of this paper, we focus on the problem of how to determine the best
partitioning for a given workload, which is the key capability that  enables
the adaptation. There are several other interesting challenges related to
adaptation, including how to efficiently manage the partitioning index and how
frequently to re-partition the disk layout. These topics are outside the scope
of this paper.





\section{Optimal Railway Design}\label{sec:optimal}  
\noindent
The optimal railway design concerns the partitioning of disk blocks into
sub-blocks such that the query I/O is minimized, while the storage overhead
induced is kept below a desired threshold. This optimization is guided by the
query workload observed by a disk block within a particular time range. Thus,
the optimization problem is localized to individual disk blocks and the sub-blocks 
created could be potentially different for different disk blocks.   

The partitioning of disk blocks into sub-blocks can be \emph{non-overlapping}
or \emph{overlapping}. In the non-overlapping case, the attributes are
partitioned among the sub-blocks with no overlap (i.e., a true partitioning). In the overlapping case, the
subset of attributes contained within sub-blocks can overlap. In both cases,
the complete graph structure for the block is replicated within the
sub-blocks, which results in a storage overhead.

In both overlapping and non-overlapping partitioning, we trade increased
storage overhead for reduced query I/O cost. In the overlapping case, the
increase in the storage overhead is higher, as some of the attributes are
replicated, in addition to the graph structure. On the other hand, enabling
overlapping attributes is expected to reduce the query I/O (in the extreme
case, there could be one sub-block per query). While the non-overlapping
partitioning scenario is a special case of the  overlapping one, specialized
algorithms can be used to solve the former problem.

In the rest of this section, we first introduce basic notation and then
formulate the overall optimization problem. The modeling of the query I/O and
storage overhead are presented next, which complete the formalization of the
optimal railway design problem.

\subsection{Basic Notation}
\noindent
Let $Q$ be the query workload, where each query $q\in Q$ accesses a set of
attributes $q.A$ and traverses parts of the graph for the time range
$q.T=[q.t_s,q.t_e]$. 
Note that when we refer to a query, we mean \emph{query kind}. That is, if
\texttt{q1} is ``all calls with a $\mbox{duration} > 100$'' and \texttt{q2} is  ``all calls
with a $\mbox{duration} > 500$'', then they are the same.
We denote the set of all attributes as $A$. Given a block
$B$, we denote its time range as $B.T$, which is the union of the time ranges
of its temporal neighbor lists. Let $s(a)$ denote the size of an attribute $a$.
We use $c_n(B)$ to denote the number of temporal neighbor lists within block
$B$ and $c_e(B)$ to denote the total number of edges in the temporal
neighbor lists within the block. We overload the notation for block size and
use $s(B)$ to denote the size of a block $B$. We have: 
\begin{equation}
s(B) = c_e(B) \cdot \Big(16 + \sum_{a\in B.A} s(a)\Big) + c_n(B) \cdot 12  
\end{equation}
Here, $16$ corresponds to the cost of storing the edge id and the timestamp,
and $12$ corresponds to the cost of storing the head vertex ($8$ bytes) plus
the number of entries ($4$ bytes) for a temporal neighbor list. 

Our goal is to create a potentially overlapping partitioning of attributes for
block $B$, resulting in a set of sub-blocks denoted by $\mathcal{P}(B)$. In
other words, we have $\bigcup_{B'\in \mathcal{P}(B)} B'.A = A$. Here,
$\mathcal{P}$ is the partitioning function.

\subsection{Optimization Problem}
\noindent
We aim to find the partitioning function $\mathcal{P}$ that minimizes the query
I/O over $B$, while keeping the storage overhead below a limit, say
$1+\alpha$ times the original. The original corresponds to the case of a single
block that contains all the attributes. Let us denote the query I/O as
$L(\mathcal{P}, B)$ and the storage overhead as $H(\mathcal{P}, B)$,
our goal is to find: 
\begin{equation} 
\mathcal{P} \leftarrow \mbox{argmin}_{\{\mathcal{P}: H(\mathcal{P}(B)) < \alpha\}} L(\mathcal{P},B)
\end{equation}

\subsection{Storage Overhead Formulation}
\noindent
The storage overhead is defined as the additional amount of disk space used to
store the sub-blocks, normalized by the original space needed by a single
block (no partitioning). The storage overhead can be formalized as follows,
for the non-overlapping case:
\begin{equation}
H(\mathcal{P}, B) = (|\mathcal{P}(B)|-1)\cdot\Big(1-\frac{c_e(B)\cdot \sum_{a\in A} s(a)}{s(B)}\Big)
\label{eq:nov-ohead}
\end{equation}

Basically, for the non-overlapping case, there is no overhead due to the
attributes, as they are not repeated. However, there is overhead for the block
structure that is repeated for each sub-block. There are $|\mathcal{P}(B)|-1$
such extra sub-blocks, and for each, the contribution to the overhead due to
storing the graph structure is given by $s(B)-c_e(B)\cdot \sum_{a\in A} s(a)$.
Eq.~\ref{eq:nov-ohead} has one nice feature, that is, it does not depend on the
details of the attribute partitioning, other than the number of partitions. We
make use of this feature, later for the ILP formulation of the problem.

For the general case of a potentially overlapping partitioning, we can
formulate the storage overhead as follows:
\begin{equation}
H(\mathcal{P}, B) = \frac{\sum_{B'\in \mathcal{P}(B)} s(B')}{s(B)} - 1 
\end{equation}

This formulation follows directly from the definition of storage overhead.
While simple, it depends on the details of the partitioning, as $s(B')$ is the
size of a sub-block $B'$, which depends on the list of attributes within the
sub-block.

\subsection{Query I/O Formulation}
\noindent
Let $m$ be a function that maps a query $q$ to the set of sub-blocks that are
accessed to satisfy it for a relevant block $B$ under a given partitioning
$\mathcal{P}$. 

For the case of non-overlapping attributes, the $m$ function lists all the 
sub-blocks whose attributes intersect with those from the query. Formally:
\begin{equation}
m(\mathcal{P}, B, q) = \{B': B'\in \mathcal{P}(B) \wedge q.A \cap B'.A \ne \emptyset\}  
\end{equation}

For the case of overlapping attributes, we use a simple heuristic to define the
set of sub-blocks used for answering the query. Algorithm~\ref{alg:greedyM}
captures it. The basic idea is to start with an empty list of sub-blocks and
greedily add new sub-blocks to it, until all query attributes are covered.  At
each iteration, the sub-block that brings the highest relative marginal gain
is  picked. The relative marginal gain is defined as  the total size of the
attributes from the sub-block that contribute to the query result, relative to
the sub-block size. While computing the marginal gain, attributes that are
already covered by sub-blocks that are selected earlier are not considered.

\begin{algorithm}[ht]
\caption{m-overlapping($\mathcal{P}, B, q$)}
\label{alg:greedyM}
\KwData{$\mathcal{P}$: partitioning function, $B$: block, $q$: query}
$S\leftarrow \emptyset; R\leftarrow \emptyset$ \tcp*{Selected attributes; Resulting sub-blocks}
\While(\tcp*[f]{While unselected attributes remain}){$S \subset q.A$}{
  $B' \leftarrow \mbox{argmax}_{B'\in\mathcal{P}(B)\setminus R} \sum_{a\in B'.A \cap q.A \setminus S} \frac{c_e(B') \cdot s(a)}{s(B')}$
  $S \leftarrow S \cup B'.A$\tcp*{Extend the selected attributes}
  $R\leftarrow R \cup B'$\tcp*{Extend the selected sub-blocks}
}
\Return R \tcp*{Final set of sub-blocks covering the query attributes}
\end{algorithm} 

Given that we have defined the function $m$ that maps a query to the set of 
sub-blocks used to answer it, we can now formalize the total query I/O cost
for a block under a given workload:
\begin{equation}
L(\mathcal{P}, B) = \sum_{q\in Q} w(q)\cdot\mathbf{1}(q.T \cap B.T \neq \emptyset) \cdot \sum_{B'\in m(\mathcal{P}, B, q)} \!\!s(B')
\end{equation}

We simply sum the I/O cost contributions of the queries to compute the total
I/O cost. A query contributes to the total I/O cost if and only if its time
range intersects with that of the block ($\mathbf{1}(q.T \cap B.T \neq
\emptyset)$). If it does, then we add the sizes of all the sub-blocks used to
answer the query to the total I/O cost. Furthermore, we multiply the I/O cost
contribution of a query with its frequency, denoted by $w(q)$ in the formula.

\section{ILP Solution}\label{sec:ilp}
\noindent
In this section, we formulate the optimal railway design problem as a mixed
Integer Linear Program (ILP). The main challenge is to represent the objective
function and the constraint as a linear combination of potentially integer
variables.

For the ILP formulation, we define a number of binary ($0$ or $1$) variables: 
\begin{itemize}
\item $x_{a,p}$: $1$ if attribute $a$ is in partition $p$, $0$ otherwise.
\item $y_{p,q}$: $1$ if partition $p$ is used by query $q$, $0$ otherwise.
\item $z_{a,p,q}$: $1$ if partition $p$ is used by query $q$ and attribute $a$
is in partition $p$, $0$ otherwise.
\item $u_{p}$: $1$ if partition $p$ is assigned at least $1$ attribute, $0$ otherwise.
\end{itemize}

Each of these variables serve a purpose:
\begin{itemize}
\item $x$s define the attribute-to-partition assignments.
\item $y$s help formulate the query I/O contribution of each partition due to
the graph structure they contain (excluding their assigned attributes).
\item $z$s help formulate the query I/O contribution of each partition, only
considering the attributes they are assigned.
\item $u$s help formulate the storage overhead requirement.
\end{itemize}

In total, we have $|A|\cdot(|A|+1)\cdot(|Q|+1)$ variables. Here, we assume that
the maximum number of partitions is fixed. In fact, we cannot have more
partitions than attributes, so the number of partitions is upper bounded by
$k=|A|$, and thus $0\leq p<k$. However, some of these partitions can be empty
in the optimal solution, which means that the number of partitions found by
the ILP solution is typically lower than the maximum possible. A simple
post-processing step removes empty partitions and creates the final
partitioning to be used. 

Finally, we define a helper notation for representing whether a variable is
accessed by a query or not: $q(a)\equiv \mathbf{1}(a \in q.A)$. 

We are now ready to state the ILP formulation. We separate the cases of
non-overlapping and overlapping partitioning, as the former case can be
formulated using smaller number of constraints.

\subsection{Non-Overlapping Partitions}\label{subsubsec:nov-ilp}
\noindent
We start with the objective function, that is the total query I/O, which is to
be minimized:
\begin{eqnarray}
\sum_{q\in Q} w(q) \cdot \Big(\sum_{p=1}^{k} \!\!&&\!\! (16\cdot c_e(B) + 12\cdot c_n(B))\cdot
y_{p,q}\nonumber\\ 
&+& \sum_{a\in A} s(a)\cdot c_e(B)\cdot z_{a,p,q}\Big)\label{eq:no-obj}
\end{eqnarray}

In Eq.~\ref{eq:no-obj}, we simply sum for each query and each partition, and
add the I/O cost of reading in the structural information found in a
sub-block, if the partition is used by the query. We then sum over each
attribute as well, and add the I/O cost of reading in the attributes. Note
that $z_{a,p,q}$ could have been replaced with $x_{a,p}\cdot y_{p,q}$, but
that would have made the objective function non-linear. 

We are now ready to state our constraints. Our first constraint is that, each
attribute must be assigned to a single partition. Formally:
\begin{eqnarray}
\forall_{a\in A}, \sum_{p=1}^{k} x_{a,p} = 1
\end{eqnarray}

Our second constraint is that, if a query $q$ contains an attribute $a$
assigned to a partition $p$, then partition $p$ is used by the query, i.e.,
$y_{p,q}=1$. In essence, we want to state: $\forall_{\{p,q\}\in [1..k]\times
Q}, y_{p,q} = \mathbf{1}(\sum_{a\in A} q(a)\cdot x_{a,p}>0)$. 

In order to formulate this constraint, we use the following ILP  construction:
Assume we have two variables, $\beta_1$ and $\beta_2$, where $\beta_2\in[0,1]$
and $\beta_1\geq 0$. We want to implement the following constraint: $\beta_2 =
\mathbf{1}(\beta_1 > 0)$. This could be expressed as a linear constraint as
follows, where $K$ is a large constant guaranteed to be larger than $\beta_1$
for all practical purposes:
\begin{eqnarray}
&& \beta_1 - \beta_2 \geq 0\nonumber\\
&& K\cdot\beta_2 - \beta_1 \geq 0\label{eq:beta-ilp}
\end{eqnarray}

We now apply this construction to our second constraint, where
$\beta_1=\sum_{a\in A} q(a)\cdot x_{a,p}$ and $\beta_2=y_{p,q}$. This results
in the following linear constraints:
\begin{eqnarray}
\forall_{\{p,q\}\in [1..k]\times Q}, 
    &&  \sum_{a\in A} q(a)\cdot x_{a,p} - y_{p,q} \geq 0 \nonumber\\
\forall_{\{p,q\}\in [1..k]\times Q}, 
    &&  K\cdot y_{p,q} - \sum_{a\in A} q(a)\cdot x_{a,p}  \geq 0 
\end{eqnarray}

Our third constraint is that, if an attribute $a$ is assigned to a partition
$p$, and partition $p$ is used by a query $q$, then the corresponding $z$
variable must be set to $1$. That is, we want: $\forall_{\{a,p,q\}\in A\times
[1..k]\times Q}, z_{a,p,q}=\mathbf{1}(x_{a,p} = y_{p,q} = 1)$. We express this
as a linear  constraint, as follows:
\begin{eqnarray}
\forall_{\{a,p,q\}\in A\times [1..k]\times Q},
    && z_{a,p,q} - (x_{a,p} + y_{p,q}) \geq -1\label{eq:no-z}
\end{eqnarray}

In Eq.~\ref{eq:no-z}, when the $x$ and $y$ variables are both $1$, the  $z$
variable is simply forced to be $1$. Otherwise, the $z$ variable can be either
$0$ or $1$, but since the $z$ variables appear in the objective function as
positive terms, the solver will set them to $0$ to minimize the I/O cost (Note
that the $z$ variables do not appear in any other constraint). 

Our fourth constraint is that, if a partition is non-empty, then its
corresponding $u$ variable must be set to $0$. In other words,  we want
$\forall_{p\in[1..k]}, u_p = \mathbf{1}(\sum_{a\in A} x_{a,p}>0)$. This is
expressed as linear constraints, as follows:
\begin{eqnarray}
\forall_{p\in[1..k]},
    && \sum_{a\in A} x_{a,p} - u_p \geq 0 \nonumber\\
\forall_{p\in[1..k]},
    && K\cdot u_p - \sum_{a\in A} x_{a,p} \geq 0 \label{eq:no-u}
\end{eqnarray}

Eq.~\ref{eq:no-u} uses the same construction as the second constraint, where
$\beta_1=\sum_{a\in A} x_{a,p}$ and $\beta_2=u_p$.

Our fifth, and the last, constraint deals with the storage overhead. We want to
 make sure that the storage overhead does not go over $\alpha$. Recall that
for  the non-overlapping attributes case, the storage overhead depends on the
number of partitions used (Eq.~\ref{eq:nov-ohead}). That means that the only 
ILP variables it depends on are the $u$s. In particular, the number of
partitions used is given by $\sum_{p=1}^{k} u_p$. This results in the
following linear constraint:
\begin{equation}
\sum_{p=1}^{k} u_p \leq 1 + \frac{\alpha}
  {1-\frac{c_e(B)\cdot \sum_{a\in A} s(a)}{s(B)}}
\end{equation}

\begin{figure}[!t]
\begin{eqnarray}
\text{minimize}  
    \sum_{q\in Q} w(q)\cdot \Big(\sum_{p=1}^{k} \!\!&&\!\! (16\cdot c_e(B) + 12\cdot c_n(B))\cdot y_{p,q}\nonumber\\
    &+& \sum_{a\in A} s(a)\cdot c_e(B)\cdot z_{a,p,q} \Big) \nonumber\\
\text{subject to}&&\nonumber\\
\forall_{a\in A}, 
    && \sum_{p=1}^{k} x_{a,p} = 1\nonumber\\
\forall_{\{p,q\}\in [1..k]\times Q}, 
    &&  \sum_{a\in A} q(a)\cdot x_{a,p} - y_{p,q} \geq 0 \nonumber\\
\forall_{\{p,q\}\in [1..k]\times Q}, 
    &&  K\cdot y_{p,q} - \sum_{a\in A} q(a)\cdot x_{a,p}  \geq 0 \nonumber\\
\forall_{\{a,p,q\}\in A\times [1..k]\times Q},
    && z_{a,p,q} - (x_{a,p} + y_{p,q}) \geq -1\nonumber\\
\forall_{p\in[1..k]},
    && \sum_{a\in A} x_{a,p} - u_p \geq 0 \nonumber\\
\forall_{p\in[1..k]},
    && K\cdot u_p - \sum_{a\in A} x_{a,p} \geq 0 \nonumber\\    
&& \sum_{p=1}^{k} u_p \leq 1 + \frac{\alpha}
  {1-\frac{c_e(B)\cdot \sum_{a\in A} s(a)}{s(B)}} \nonumber
\end{eqnarray}
\caption{ILP formulation for the non-overlapping optimal railway design.}
\label{fig:nov-ilp}
\end{figure}

The final ILP formulation for the non-overlapping partitioning is given in
Figure~\ref{fig:nov-ilp}. We have a total of $|A|^2\cdot|Q| +
2\cdot|A|\cdot|Q| + 3\cdot|A| + 1$ constraints and the objective function
contains $|A|\cdot|Q|\cdot(1+|A|)$ variables.

\subsection{Overlapping Partitions}\label{subsubsec:ov-ilp}
\noindent
We present an ILP formulation of the problem, as we did for the case of
non-overlapping partitions in Section~\ref{subsubsec:nov-ilp}. We use the same
set of variables and the same objective function. However, the formulation of
the constraints differ. 

Our first constraint is that, each attribute must be assigned to at least one
partition. Formally:
\begin{eqnarray}
\forall_{a\in A}, \sum_{p=1}^{k} x_{a,p} \geq 1
\end{eqnarray}

As our second constraint, we require that for each attribute contained in a
query, there needs to be a partition that is used by that query and that
contains the attribute in question. Formally:
\begin{eqnarray}
\forall_{\{a,q\}\in A\times Q}, \sum_{p=1}^{k} z_{a,p,q} \geq q(a) 
\end{eqnarray}

\begin{sloppypar}
As our third constraint, we require that if a query is using an attribute from
a partition, then that partition must contain the attribute. I.e., we need
to link the $z$ variables with the $x$ variables as
$\forall_{\{a,p,q\}\in A\times [1..k]\times Q}, (z_{a,p,q} = 1) \implies 
(x_{a,p} = 1)$. This can be stated as linear constraints:
\begin{eqnarray}
\forall_{\{a,p,q\}\in A\times [1..k]\times Q}, x_{a,p} - z_{a,p,q} \geq 0 
\end{eqnarray}
\end{sloppypar}

As our fourth constraint, we require that if a query is using at least one
attribute from a partition, then that partition must be used by the query.
I.e., we need to link the $z$ variables with the $y$ variables as
$\forall_{\{p,q\}\in [1..k]\times Q}, y_{p,q} = \mathbf{1}(\sum_{a\in A}
z_{a,p,q}>0)$. As before, we use the ILP construction from
Eq.~\ref{eq:beta-ilp} for this, where $\beta_2=y_{p,q}$ and $\beta_1 =
\sum_{a\in A} z_{a,p,q}$. We get:
\begin{eqnarray}
\forall_{\{p,q\}\in [1..k]\times Q}, 
    &&  \sum_{a\in A} z_{a,p,q} - y_{p,q} \geq 0 \nonumber\\
\forall_{\{p,q\}\in [1..k]\times Q}, 
    &&  K\cdot y_{p,q} - \sum_{a\in A} z_{a,p,q} \geq 0 
\end{eqnarray}

Our fifth constraint is that, if an attribute $a$ is assigned to a partition
$p$, and partition $p$ is used by a query $q$, then the corresponding
$z_{a,p,q}$ variable must be set to $1$. This is same as the formulation for
the non-overlapping case from Eq.~\ref{eq:no-z}.

Our sixth constraint is that, if a partition is non-empty, then its
corresponding $u$ variable must be set to $0$. Again, this is same as the
formulation for the non-overlapping case from Eq.~\ref{eq:no-u}.

Our seventh, and the last, constraint deals with the storage overhead. However,
the storage overhead formulation for the overlapping case is different from 
the one for the non-overlapping case. This is because the overhead does not
merely depend on the number of partitions, as attributes might have to be read
multiple times from different partitions (due to the overlaps). As a result,
we express the overhead using base variables as in the objective function.
Formally:
\begin{eqnarray}
\sum_{p=1}^{k} \Big((16\cdot c_e(B) &+& 12 \cdot c_n(B)) \cdot u_p  \nonumber \\ 
+ \sum_{a\in A} s(a) \!\!&\cdot&\!\! c_e(B)\cdot x_{a,p} \Big) \leq s(B)\cdot (1+\alpha)
\end{eqnarray}

\begin{figure}[!t]
\begin{eqnarray}
\text{minimize}  
    \sum_{q\in Q} w(q)\cdot \Big(\sum_{p=1}^{k} \!\!&&\!\! (16\cdot c_e(B) + 12\cdot c_n(B))\cdot y_{p,q}\nonumber\\
    &+& \sum_{a\in A} s(a)\cdot c_e(B)\cdot z_{a,p,q} \Big) \nonumber\\
\text{subject to}&&\nonumber\\
\forall_{a\in A}, 
    && \sum_{p=1}^{k} x_{a,p} \geq 1\nonumber\\
\forall_{\{a,q\}\in A\times Q},
    &&  \sum_{p=1}^{k} z_{a,p,q} \geq q(a) \nonumber\\
\forall_{\{a,p,q\}\in A\times [1..k]\times Q}, 
    && x_{a,p} - z_{a,p,q} \geq 0 \nonumber\\
\forall_{\{p,q\}\in [1..k]\times Q}, 
    &&  \sum_{a\in A} z_{a,p,q} - y_{p,q} \geq 0 \nonumber\\
\forall_{\{p,q\}\in [1..k]\times Q}, 
    &&  K\cdot y_{p,q} - \sum_{a\in A} z_{a,p,q}  \geq 0 \nonumber\\
\forall_{\{a,p,q\}\in A\times [1..k]\times Q}, 
    && z_{a,p,q} - (x_{a,p} + y_{p,q}) \geq -1 \nonumber\\
\forall_{p\in[1..k]},
    && \sum_{a\in A} x_{a,p} - u_p \geq 0 \nonumber\\
\forall_{p\in[1..k]},
    && K\cdot u_p - \sum_{a\in A} x_{a,p} \geq 0 \nonumber\\    
\sum_{p=1}^{k} \Big((16\cdot c_e(B) &+& 12 \cdot c_n(B)) \cdot u_p  \nonumber \\ 
+ \sum_{a\in A} s(a) \!\!&\cdot&\!\! c_e(B)\cdot x_{a,p}  \Big)\leq s(B)\cdot (1+\alpha)\nonumber
\end{eqnarray}
\caption{ILP formulation for the overlapping partitioning}
\label{fig:ov-ilp}
\end{figure}

The final ILP formulation for the overlapping partitioning is given in
Figure~\ref{fig:ov-ilp}. We have a total of $2\cdot|A|^2\cdot|Q| +
3\cdot|A|\cdot|Q| + 3\cdot|A| + 1$ constraints and the objective function
contains $|A|\cdot|Q|\cdot(1+|A|)$ variables.

\section{Heuristic Solution}\label{sec:heuristic}
\noindent
The ILP formulation described in Section~\ref{sec:optimal} finds an optimal
solution to the problem of partitioning disk blocks into sub-blocks such that
the query I/O is minimized. Unfortunately, solving these types of constraint
problems at scale can become a performance bottleneck, since integer
programming is NP-Hard. In a graph database using the railway layout, the 
layout optimization of a block should be fast enough so that it could be
piggybacked on disk I/O when significant workload change that necessitates a
new layout is detected. We therefore introduce heuristic algorithms for both
overlapping and non-overlapping partitioning scenarios.  Experiments in
Section~\ref{sec:evaluation} demonstrate that these heuristic algorithms show
significantly improved running times over the optimal approaches, while still
appreciably reducing the query I/O cost.

\begin{algorithm}[t!]
\caption{Algorithm for partitioning blocks into sub-blocks with non-overlapping attributes.}
\label{alg:non-overlappingP}
\KwData{$B$: block, $Q$: set of queries}
$c^*\leftarrow \infty$ \tcp*{Lowest cost over all \# of partitions}
\For(\tcp*[f]{For each possible \# of partitions}){$k=1$ to $|A|$}{
   $R[i]\leftarrow \emptyset, \forall i\in [1..k]$ \tcp*{Initialize partitions}
   \For(\tcp*[f]{For each attribute}){$a \in A$\textnormal{, in decr.\/ order of }$f(a)$}{
      $c\leftarrow \infty$ \tcp*{Lowest cost over all assignments}  
      $j\leftarrow -1$ \tcp*{Best partition assignment}
      \For(\tcp*[f]{For each partition assignment}){$i\in [1..k]$} {
         $R[i]\leftarrow R[i] \cup \{a\}$\tcp*{Assign attribute}
         \If(\tcp*[f]{If query cost is lower}){$L(R, B, Q)<c$}{
            $c\leftarrow L(R, B, Q)$\tcp*{Update the lowest cost}
            $j\leftarrow i$\tcp*{Update the best partition}
         }
         $R[i]\leftarrow R[i] \setminus \{a\}$\tcp*{Un-assign attribute}
      }
      $R[j]\leftarrow R[j] \cup \{a\}$\tcp*{Assign to best partition}
   }
   \lIf(\tcp*[f]{If solution infeasible}){$H(R, B, Q)>\alpha$}{\textbf{break}}
   \If(\tcp*[f]{If solution has lower cost}){$L(R, B, Q)<c^*$}{
     $c^* \leftarrow L(R, B, Q)$\tcp*{Update the lowest cost}
     $\mathcal{P}(B)\leftarrow R$\tcp*{Update the best partitioning}
   } 
}
\Return $\mathcal{P}(B)$ \tcp*{Final set of sub-blocks}
\end{algorithm} 

\subsection{Non-Overlapping Attributes}\label{subsec:nov-heuristic}
\noindent
For the non-overlapping attributes scenario, we use a heuristic algorithm that
greedily assigns attributes to partitions. The pseudo-code of it is given in
Algorithm~\ref{alg:non-overlappingP}. One complication is that, the number of
partitions is not known a priori. Yet, we know that the number of partitions
is bounded by the number of attributes. As such, we start with a single
partition, and try different number of partitions, until we hit the maximum
number of partitions or the storage overhead goes beyond the threshold
$\alpha$.  Among all partition counts tried, the one that provides the
lowest query cost is selected as the final partitioning. Note that, for the
non-overlapping scenario, the storage overhead is an increasing function of
the number of partitions. As such, once we exceed the storage overhead
threshold, we can safely stop trying larger numbers of partitions. 

For a fixed number of partitions, the algorithm operates by incrementally
assigning attributes to partitions. We consider the attributes in decreasing
order of their frequency. This is because the reverse, that is assigning
highly frequent attributes later, may result in making assignments that are
hard to  balance out later. Initially, all partitions are empty. We pick the
next unassigned attribute and evaluate assigning it to one of the available
partitions. The assignment that results in the lowest query cost is selected
as the best assignment and is applied. When computing the query cost, we only
consider the attributes assigned so far.

\paragraph*{Computational Complexity} The computational complexity of the
algorithm is $\mathcal{O}(k^2\cdot |A|\cdot |Q|)$, where $k$ is the maximum 
number of partitions tried. The $|Q|$ term is the number of unique queries and
comes from the  cost of computing the query I/O (this can be computed
incrementally, even though this is not shown in the pseudo-code). While in the
worst case we have $k=|A|$, resulting in a computational complexity of
$\mathcal{O}(|A|^3\cdot |Q|)$, in practice $k$ is much lower due to the upper
bound $\alpha$ on the storage overhead.

\begin{algorithm}[t!]
\caption{Algorithm for partitioning blocks into sub-blocks with overlapping attributes.}
\label{alg:overlappingP}
\KwData{$B$: block, $Q$: set of queries}
$\mathcal{P}(B) \leftarrow \{q.A: q\in Q\}$ \tcp*{Each query gets its own sub-block}
$A'\leftarrow A \setminus \bigcup_{q\in Q} q.A$ \tcp*{Attributes not covered by the queries}
\If(\tcp*[f]{There are uncovered attributes}){$A'\neq \emptyset$}{
  $\mathcal{P}(B) \leftarrow \mathcal{P}(B) \cup \{A'\}$\tcp*{Add missing attributes}
}
\While(\tcp*[f]{Until storage overhead is below $\alpha$}){$H(\mathcal{P},B) > \alpha$}{
  $c^{*}\gets \infty $ \tcp*{Lowest cost over all sub-block pairs}
  $(b_x,b_y)\gets (\emptyset,\emptyset)$ \tcp*{Sub-block pair with the lowest cost}
  \For(\tcp*[f]{For each pair of blocks}){$\{b_i,b_j\}\in\mathcal{P}(B)$}{
    $\mathcal{P'}(B) \leftarrow \mathcal{P}(B) \setminus \{b_i, b_j\} \cup \{b_i \cup b_j\}$\\
    $c\gets \frac{L(\mathcal{P}',B,Q)-L(\mathcal{P},B,Q)}{H(\mathcal{P},B)-H(\mathcal{P}',B)}$\tcp*{Cost of merge}
    \If(\tcp*[f]{Cost is lower}){$c<c^{*}$}{
        $c^{*}\gets c$\tcp*{Update the lowest cost}
        $(b_x,b_y)\gets (b_i,b_j)$\tcp*{Update the best pair}
    }
  }
  $\mathcal{P}(B) \leftarrow \mathcal{P}(B) \setminus \{b_x, b_y\} \cup \{b_x \cup b_y\}$ \\
}
\Return $ \mathcal{P}(B)$  \tcp*{Final set of sub-blocks}
\end{algorithm} 
\subsection{Overlapping Partitions}\label{subsec:ov-heuristic}
\noindent 
For the overlapping attributes scenario, we use a heuristic algorithm that
starts with each query in its own partition and greedily merges partitions
until the storage overhead is below the limit. The pseudo-code of it is given
in Algorithm~\ref{alg:overlappingP}.

We start the algorithm in a state where for each unique query there is a
separate sub-block that contains the attributes from that query. If there are 
attributes not covered by the queries, they are assigned to a special
sub-block. This is the ``ideal'' partitioning, because the I/O cost would be
minimized for the workload at hand. However, in most practical settings, this
partitioning will have excessive storage overhead. Thus, we iteratively
combine the pair of partitions that has the lowest cost. This is repeated
until the storage overhead is below the threshold $\alpha$. The end result is
the final overlapping partitioning. 

We define the cost of a merge based on the query I/O and storage cost. In
particular, we measure the increase in the query I/O due to the merge, per
reduction in the  storage space used. We want to minimize this metric. More formally, assuming
$\mathcal{P}$ is the partitioning before the merge and $\mathcal{P}^\prime$ is
the partitioning after the merge, the utility can be formulated as:
$$
\frac{L(\mathcal{P}',B,Q)-L(\mathcal{P},B,Q)}{H(\mathcal{P},B)-H(\mathcal{P}',B)}
$$

\paragraph*{Computational Complexity} The computational complexity of the
algorithm is $\mathcal{O}(|A|\cdot |Q|^3)$. At each iteration, the algorithm
reduces the number of partitions by one and initially there are $|Q|$
partitions. As such, in the worst case, there will be $|Q|$ iterations. The
number of pairs considered is bounded by $|Q|^2$. The utility metric can be
computed incrementally, but requires iterating over the query attributes,
bringing in the $|A|$ term.

\section{Evaluation}\label{sec:evaluation}
\noindent
In this section, we describe our prototype implementation, and the results of
our evaluation, demonstrating that the railway layout scheme significantly
reduces query I/O for interaction graphs.

\begin{figure*}[ht!]
\centerline{\begin{tabular}{c@{}c@{}c}
\includegraphics[width=0.33\textwidth]{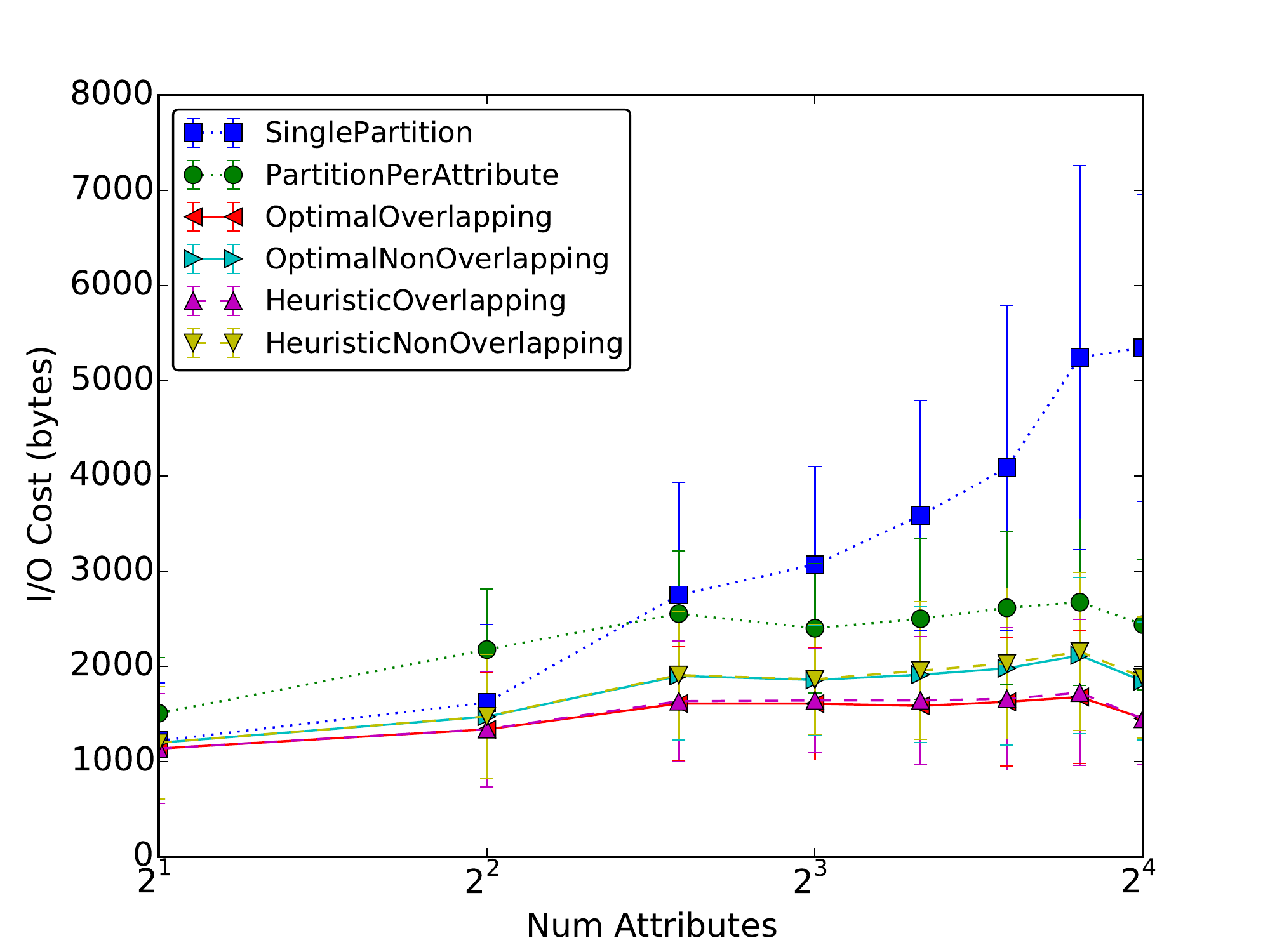} &
\includegraphics[width=0.33\textwidth]{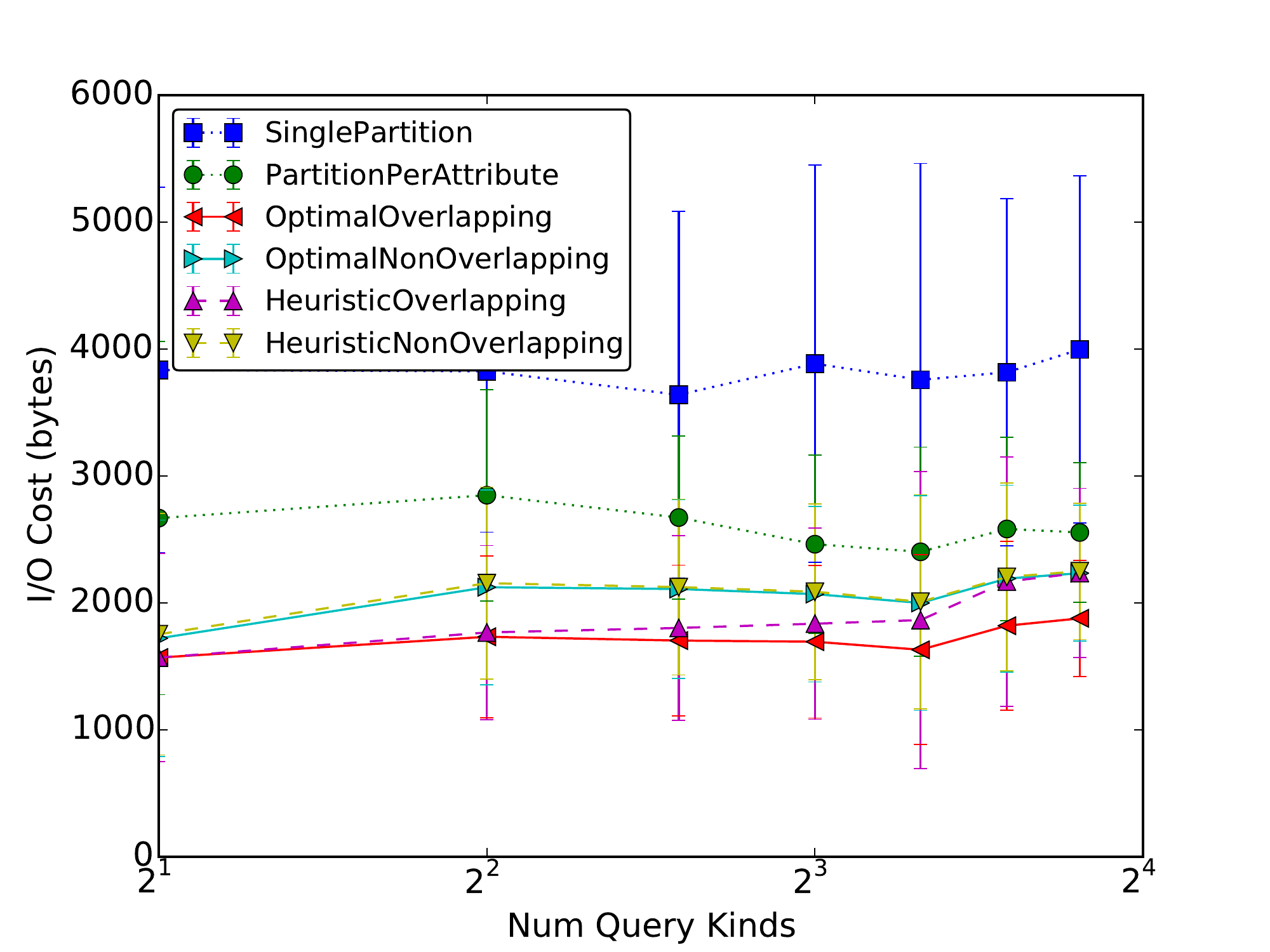} &
\includegraphics[width=0.33\textwidth]{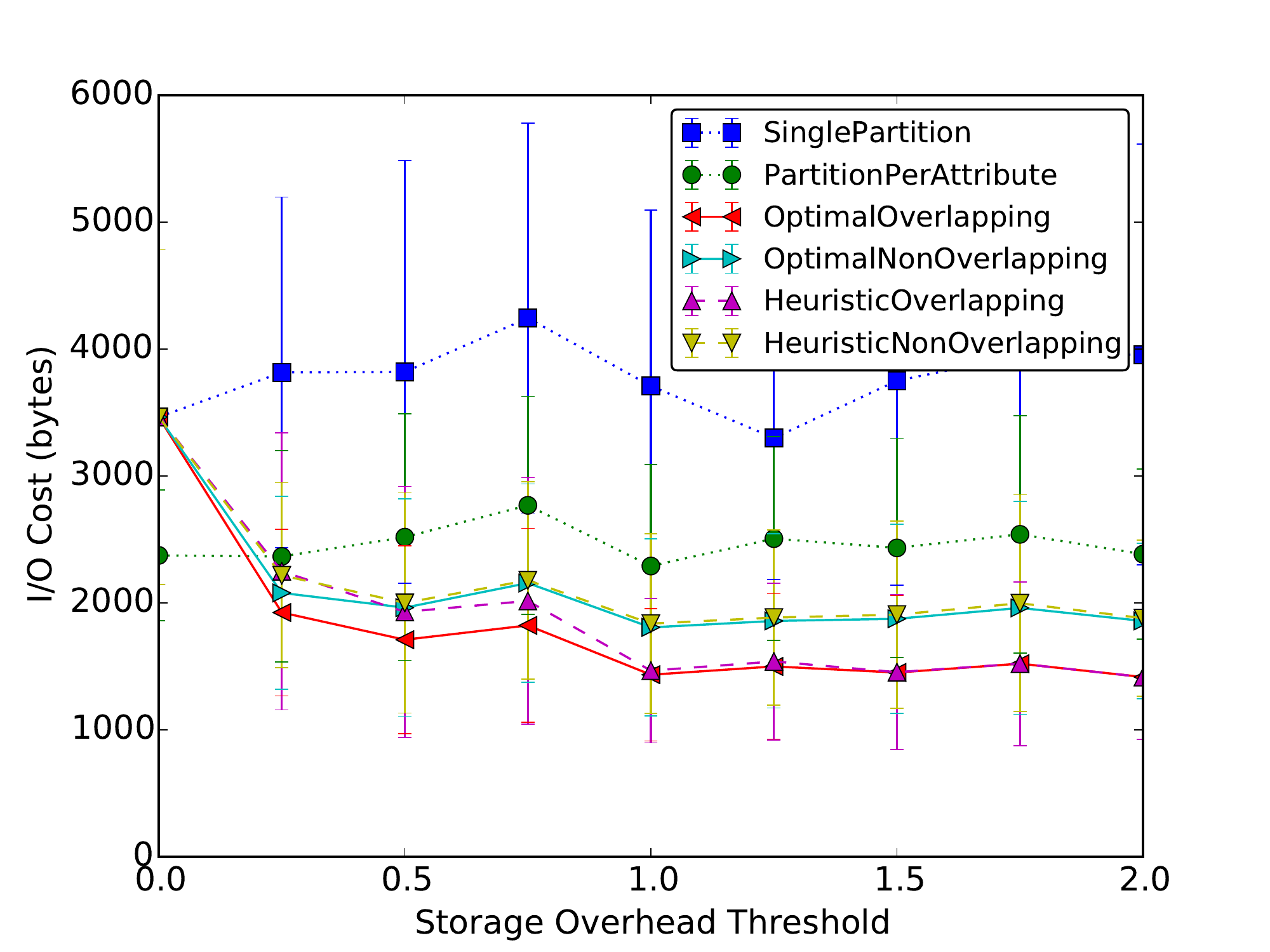}\\
\end{tabular}}
 \caption{Query I/O cost for different partitioning algorithms for increasing
   number of attributes, number of query kinds, and for increasing storage
   overhead threshold.}
 \label{fig:queryio}
 \end{figure*}

\begin{figure*}[ht!]
\centerline{\begin{tabular}{c@{}c@{}c}
\includegraphics[width=0.33\textwidth]{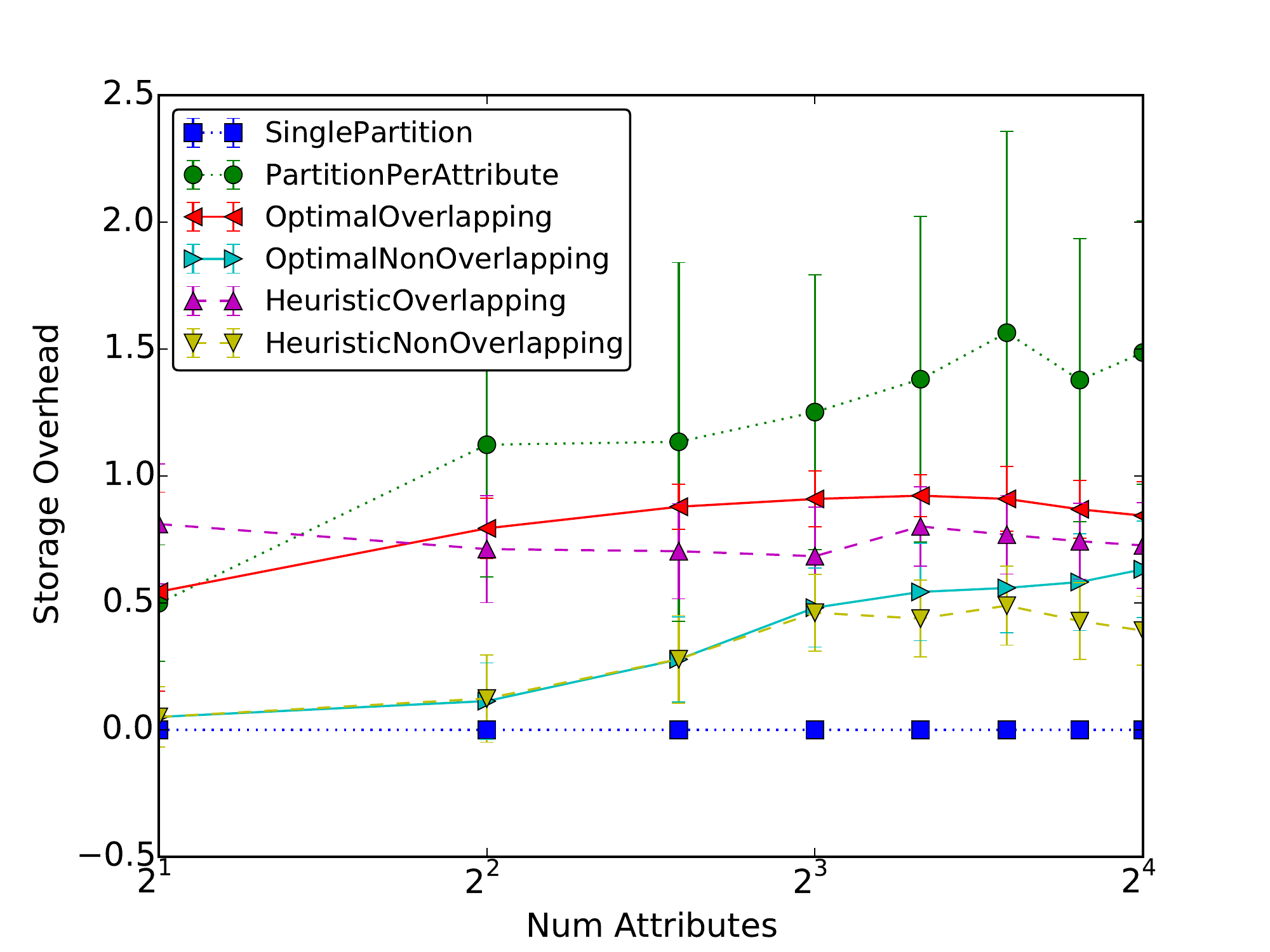} &
\includegraphics[width=0.33\textwidth]{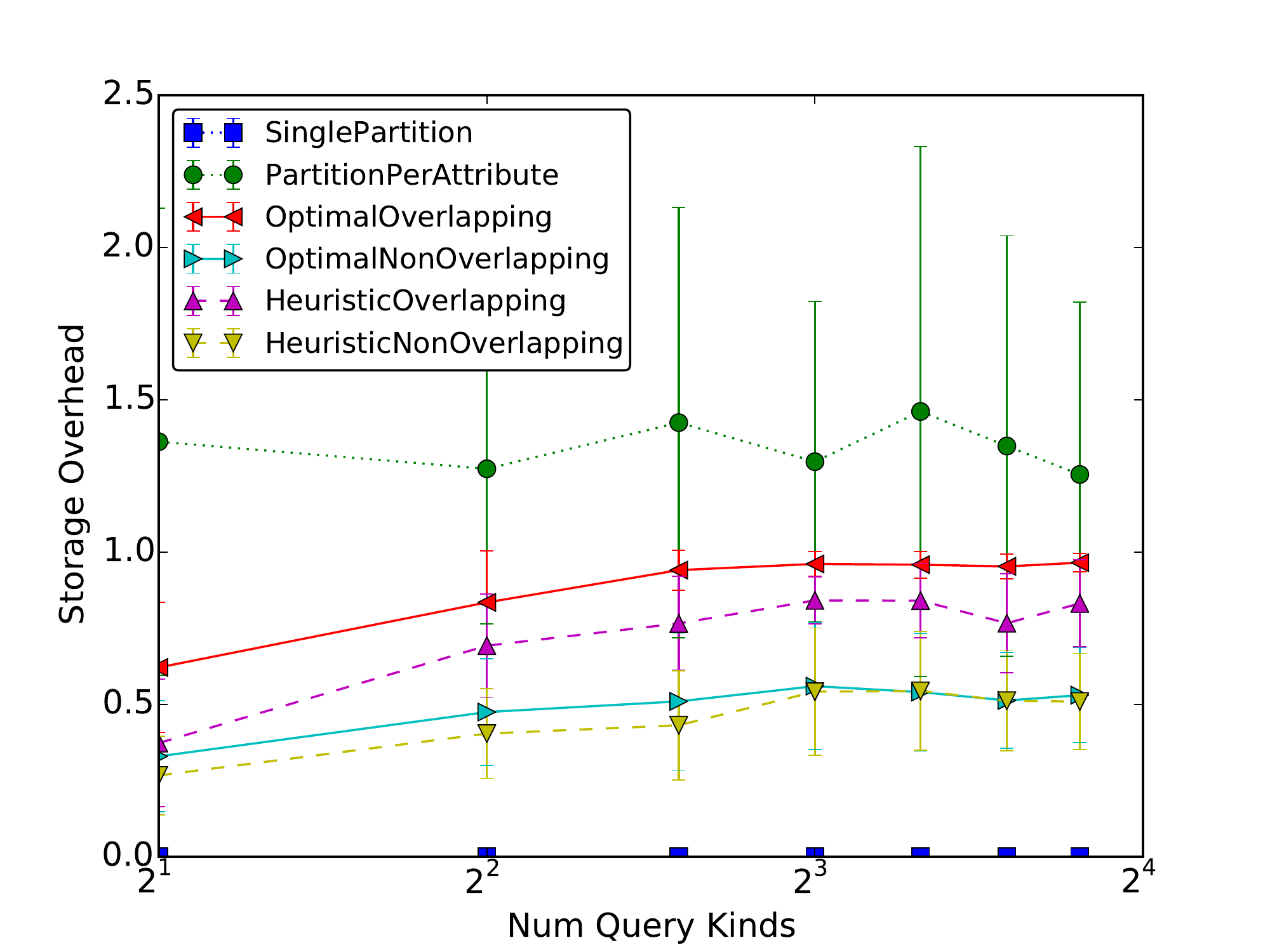} &
\includegraphics[width=0.33\textwidth]{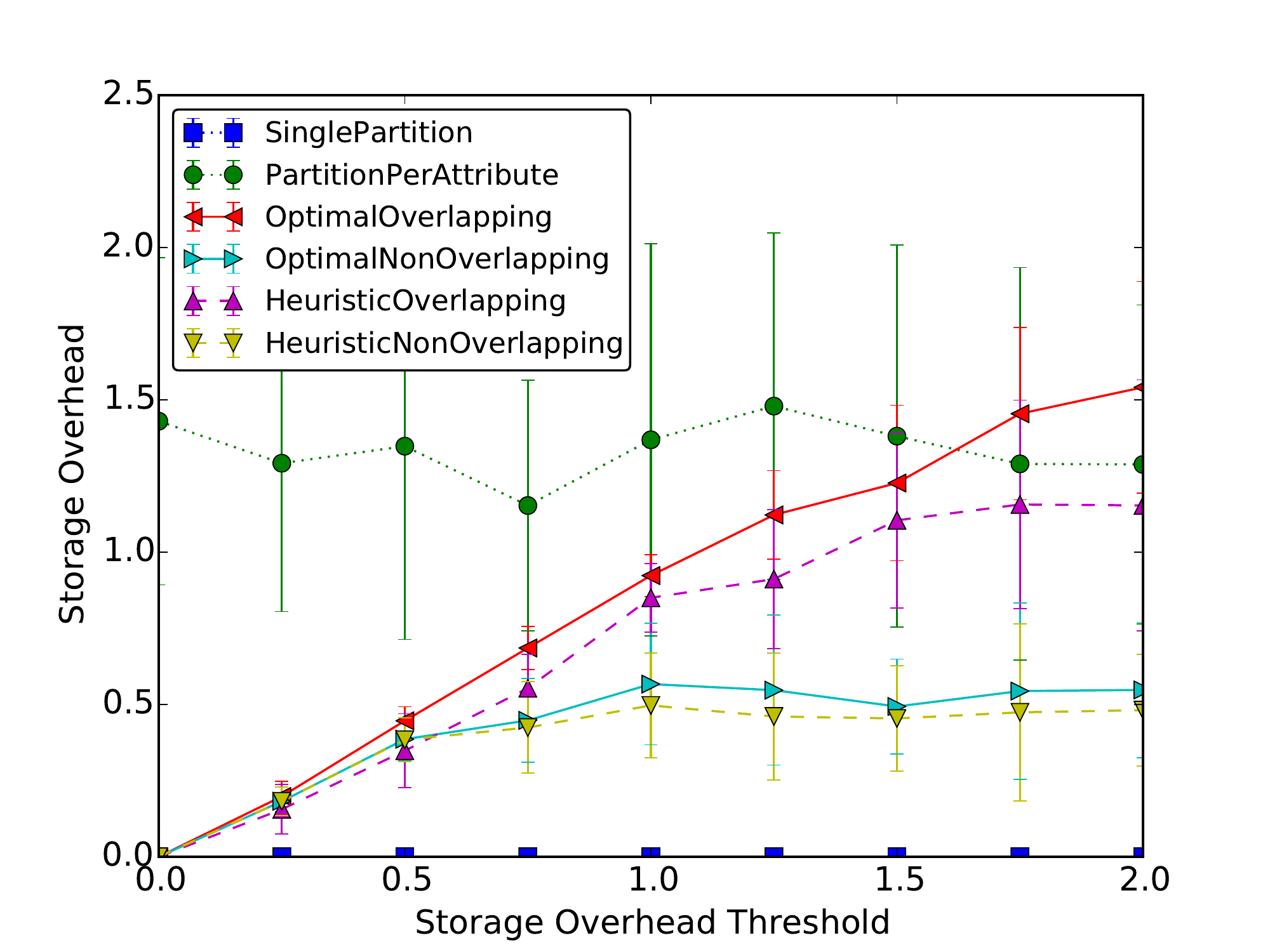}\\
\end{tabular}}
 \caption{Storage overhead for different partitioning algorithms for increasing
   number of attributes, number of query kinds, and for increasing storage
   overhead threshold.}
 \label{fig:storage}
 \end{figure*}

\begin{figure*}[ht!]
\centerline{\begin{tabular}{c@{}c@{}c}
\includegraphics[width=0.33\textwidth]{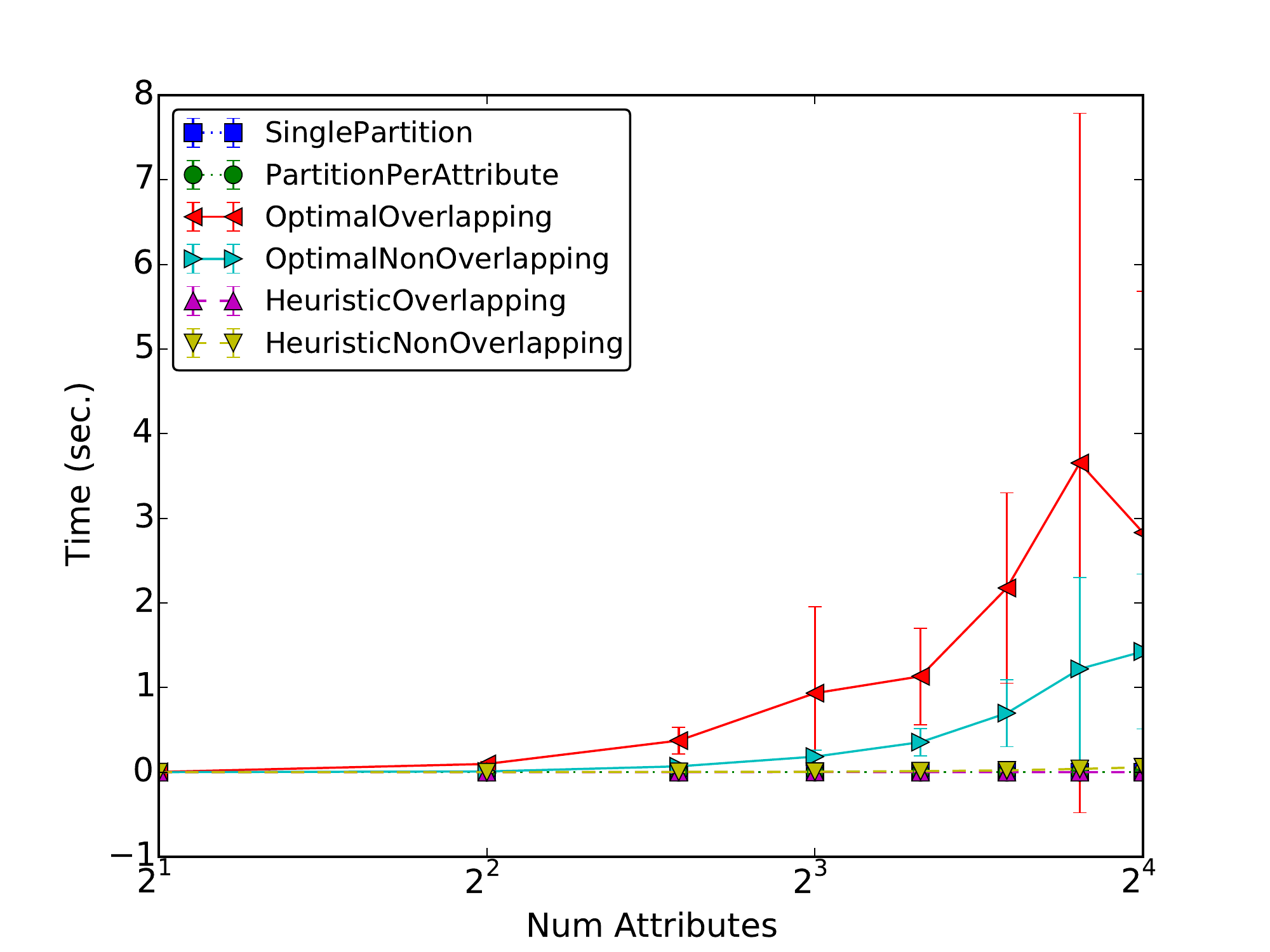} &
\includegraphics[width=0.33\textwidth]{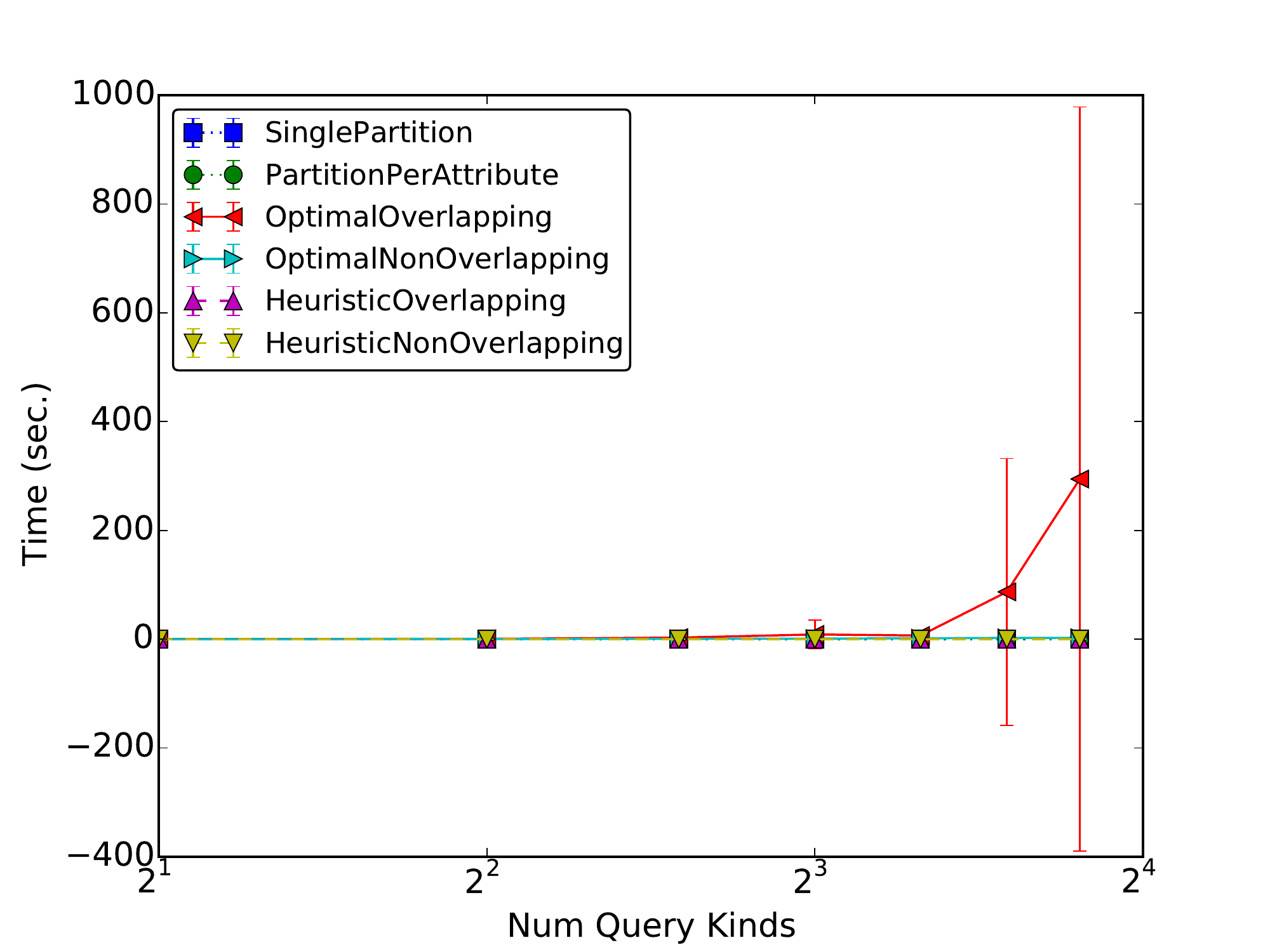} &
\includegraphics[width=0.33\textwidth]{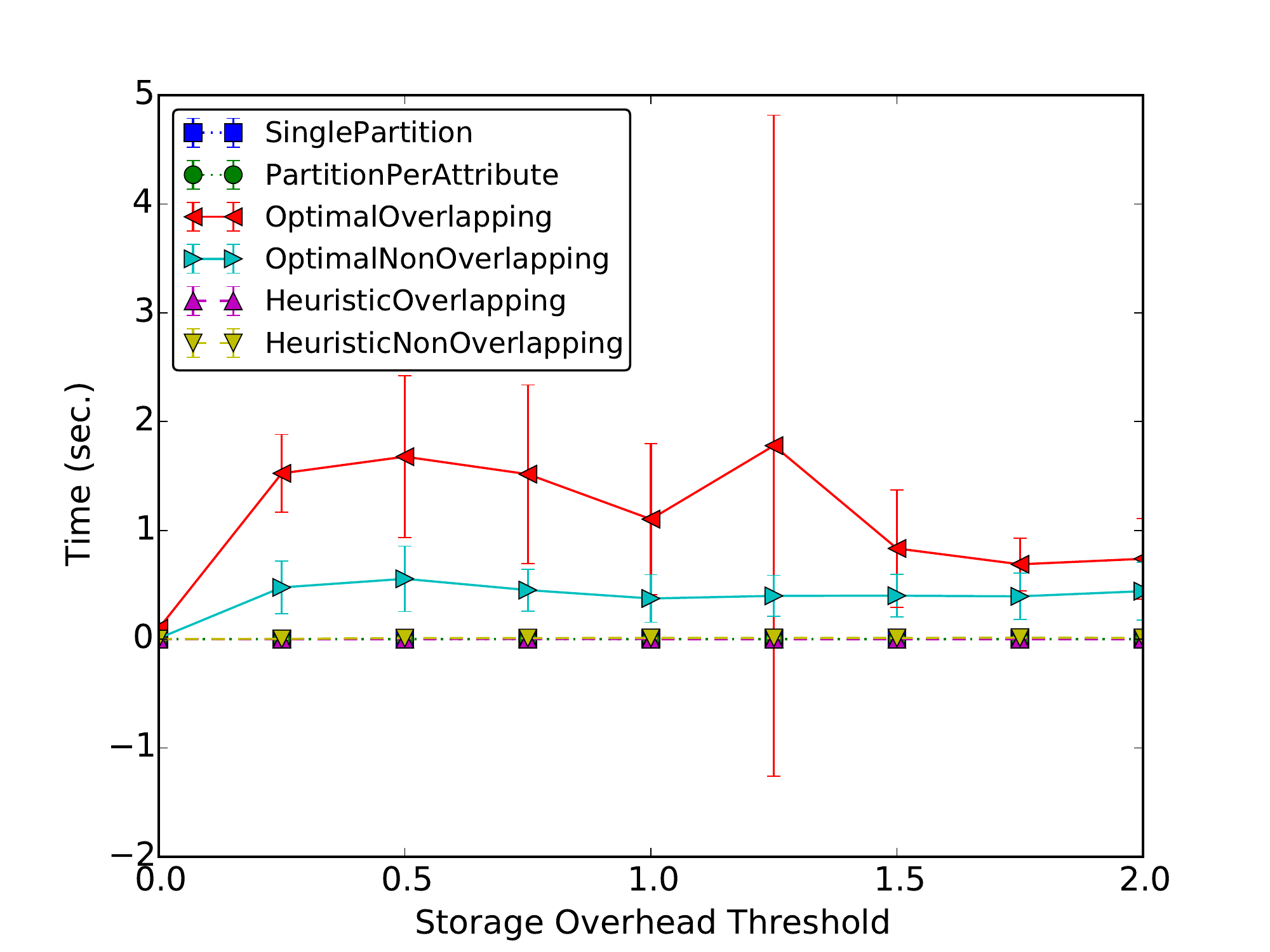}\\
\end{tabular}}
 \caption{Running time of different partitioning algorithms for increasing
   number of attributes, number of query kinds, and for increasing storage
   overhead threshold.}
 \label{fig:scalability}
 \end{figure*}

\subsection{Implementation}
\noindent
We have implemented the four partitioning algorithms described in this paper,
and a workload simulator to evaluate the algorithms and our disk layout
design. All source code for the algorithms, as well as the simulator and
experiments are publicly
available\footnote{https://github.com/usi-systems/graphdb}.

\paragraph*{Algorithms}
The partitioning algorithms were written in C++ using the LLVM 3.5 compiler. The
implementation uses data structures from the graph database implementation
described in our prior work~\cite{gedik14} to represent queries and
disk-blocks. To solve the ILP formulation of the partitioning problem, it relies
on the C libraries from the Gurobi Optimizer~\cite{gurobi} software.

\begin{table}
\centering
\begin{tabular}{|l||c|c|}
\hline
Parameter         & Default                                          \\ \hline\hline
\# of attributes  & $10$                                             \\ \hline
attribute sizes   & Zipf($z=0.5$, $\{4, 1, 8, 2, 16, 32, 64\}$) \\ \hline
query length      & Normal($\mu=3$, $\sigma=2.0$)                    \\ \hline
\# of query kinds & $5$                                              \\ \hline
query kind freq.\/ & Zipf($z=0.5$, $n=5$)                       \\ \hline
storage ohd. threshold &$\alpha=1.0$                       \\ \hline
\end{tabular}
\caption{Workload generation parameter defaults}\label{tbl:workload-params}
\end{table}
\paragraph*{Workload simulator}
To evaluate our design, we also implemented a workload simulator, whose
parameters and default settings are given in Table~\ref{tbl:workload-params}.
The default number of attributes in the graph database schema is taken as
$10$, even though we experiment with a range of values for it. The size of the
attributes come from  the list of sizes given in
Table~\ref{tbl:workload-params} and are picked randomly from a Zipf
distribution with $z=0.5$. The average number of unique query types we
have in the workload for a particular time point is taken as $5$, which is
another parameter we vary throughout our experiments. The frequencies of
different queries follow a Zipf distribution with $z=0.5$.

\subsection{Environment}
\noindent
We ran all experiments on a machine with a 2.3 GHz Intel i7 processor that has
32 KB L1 data, 32 KB L1 instruction, 256 KB L2 (per core), 6 MB L3 (shared)
cache, and 16 GB of main memory. The processor has four cores, but our
implementation only uses a single core. The operating system was OS X 10.9.4.

\subsection{Experiments} 
\noindent
Our experiments evaluate three aspects of our design:
(i) the reduction in query I/O due to using the railway layout,
(ii) the expected increase in storage cost resulting from the railway layout, 
and (iii) the scalability of the partitioning algorithms

We measured these three respective values, \emph{query I/O cost},
\emph{storage overhead}, and \emph{running time}, for each partitioning
algorithm, as we varied three parameters to the default workload in
Table~\ref{tbl:workload-params}:
\begin{itemize}
\item \emph{Number of attributes} is the total number of attributes 
in the iteration graph schema. We increased the attribute count by multiples of two from 2 to 16.
\item \emph{Number of query kinds} is the number of unique query types in the
  workload. We increased the number of query types by multiples of two from 2 to
  14. Beyond 14, the optimal solvers were no longer able to find solutions in a
  reasonable amount of time.
\item \emph{Storage overhead threshold} is the user-specified parameter that
dictates how much storage overhead will be tolerated for a solution. 
We increased the storage overhead threshold by increments of 0.25 from 0 to 2.0.
\end{itemize}

For all experiments, other than the experiment in which we explicitly altered
the value, we used a default storage overhead threshold value of 1.0. We believe
this is a reasonable number, as it corresponds with doubling the available
storage space.

 As baseline comparisons, we also measured the results for two na\"{\i}ve
partitioning schemes:  SinglePartition places all attributes into a single
partition, and PartitionPerAttribute creates a separate partition for each
attribute. The SinglePartition scheme represents the standard disk
layout, and the PartitionPerAttribute approach represents an extreme partitioning
(although not an optimal one, as it potentially increases both the query I/O and
storage costs).

For each configuration, we ran the experiment 10 times.  Each partitioning
algorithm used the same workload for each run, but each run was on a different
random workload using the same configuration parameters. We report the average
(arithmetic mean) and standard deviation.

\paragraph*{Query I/O}
Figure~\ref{fig:queryio} shows the results from the query I/O cost
measurements. In all three experiments, we see the benefit of the railway
layout. The SinglePartition and PartitionPerAttribute layouts represent baseline
measurements for a traditional layout and pathological partitioning scheme. All
versions of the railway layout result in better query I/O than the baseline
measurements, except when the storage threshold is set to not allow any
overhead (as we would expect).

In the left graph, we see that the benefits of the railway layout become more
pronounced as we increase the number of attributes. At the low end of the graph,
with a schema of only 2 attributes, the optimal overlapping partitioning
algorithm results in a 7 percent reduction in query I/O cost. However, at the
high end, with 16 attributes, there is a 73 percent reduction in I/O cost. Note
that the heuristic overlapping is almost as good, giving a 72 percent reduction
in I/O cost.

In the middle figure, we see that the benefits of the railway layout remain
relatively constant as we increase the number of query kinds. 
In the 2 query case, we see a 59 percent difference between the  optimal
overlapping and single partitioning schemes, while at the 14 query case, we see
 a 53 percent difference. While increasing the number of query kinds did not
 have a big impact on query I/O, it did have a large impact on running time, as
 we will see.

The railway layout makes a tradeoff between query I/O cost and storage cost. We
see in the right graph of Figure~\ref{fig:queryio}, when the user explicitly
disallows any increase in storage (i.e., sets the threshold to 0), then the
railway layout does not help. However, with even just a slight 25 percent
increase in storage, all railway layouts reduce query I/O, demonstrating
reductions of 45 percent.

\paragraph*{Storage Overhead}
The experiments in Figure~\ref{fig:storage} quantify the storage overhead that
one can expect with using the railway layout. In the left graph, we see that the
optimal overlapping and heuristic overlapping approach the user specified limit
of doubling the storage space. As expected, the algorithms will make use of
extra storage in order to reduce the query I/O cost. The non-overlapping schemes
are limited in the amount or storage overhead that they use, since they cannot
duplicate attributes in separate partitions. So, the extra storage overhead is
attributed to duplicating the graph structure.

The middle graph shows a similar result. The overlapping partitioning algorithms
approach the user specified threshold, while the non-overlapping schemes are
bounded.

The right graph in Figure~\ref{fig:storage} is interesting. It shows that as
the user increases the threshold to a value of 2.0 (i.e., tripling the available
storage) both optimal schemes will try to take advantage of the extra space to
reduce query I/O.

\paragraph*{Scalability}
The experiments in Figure~\ref{fig:scalability} show the running times for our four algorithms.
As we can see in the left graph, when the schema had 14 attributes. The optimal
overlapping scheme took 3.64 seconds to find a solution, and the optimal
non-overlapping took 1.22 seconds. In contrast, both heuristic solutions took
deciseconds (i.e., 1/10s of a second) to solve.

The number of query kinds had a large impact on solving time. After leaving the
experiment running for more than 12 hours, we were not able to complete the optimal overlapping
measurement for the case of 16 query kinds. This experiment demonstrates the
benefit of our heuristic greedy algorithms.

However, as shown in the right graph, the storage overhead threshold did not
have a significant impact on the running time. This is as expected, since the
optimal solvers scale with the number of variables in the constraint problem,
and the number of variables does not increase as we alter the storage overhead threshold.

\paragraph*{Summary}
Overall, our experiments demonstrate the benefits of the railway layout.  For a
storage increase of just 25\%, the optimal partitioning algorithm reduces the
query I/O cost by 45\%.  When allowed to double the storage usage, the
overlapping partitioning algorithm can reduce the I/O cost by 73\%. The
heuristic algorithm performs almost as well, reducing the I/O cost by 72\%, while
also reducing the running time needed to find a solution by orders of magnitude.

\section{Related Work}\label{sec:related}
\noindent
There has recently been increased research interest in large-scale graph
analysis and programming models. These include synchronous vertex programming
pioneered by Pregel~\cite{malewicz10}, such as Apache Giraph~\cite{giraph};
asychronous vertex programming pioneered by GraphLab~\cite{gonzalez12,kyrola12},
and generalized iterated matrix-vector multiplication pioneered by
PEGASUS~\cite{kang09}. These systems largely focus on the problem on analytical
processing, while our work focuses on data management. Moreover, the graphs
these systems provide do not have a temporal dimension.

The railway layout and algorithms build on our prior work~\cite{gedik14}, which
added a temporal dimension to the notion of locality for organizing the disk
layout of interaction graph databases. Graph database nodes are placed in the
same disk block if they are close together both spatially and temporally. The
railway layout extends this design to partition disk blocks into sub-blocks that
reduces the query I/O cost. Because interaction graphs are append only, the
railway design enables the disk layout to adapt with changing workloads.

Our adaptation scheme is similar to work on adaptive layouts for relational
database. The $H_2O$~\cite{alagiannis14} system can adapt its data layout into
three types, row-major, column-major, or groups of columns, depending on the
workload. HYRISE~\cite{grund10} provides a similar adaptive layout scheme for an
in-memory relational database. Both systems use heuristic, iterative solutions
to determine partitioning. The railway layout scheme differs, in that it
targets interaction graphs, and we present optimal solutions, in addition to 
heuristic solutions.

The rise in popularity of social networks, and the recognition that workloads
for social network data differ from traditional workloads, has lead to increased
scrutiny on the problem of disk layout for graph databases.
Bondhu~\cite{hoque12}, the layout manager for the Neo4j graph
database~\cite{neo4j}, aims to minimize the number of seek operations for
small user block sizes by fetching multiple friends' data at the same time, and
by clustering related data into the same block. Bondhu differs from our work in
that the cost model does not include a notion of time, nor does it allow for
adaptive layouts.

Instead of storing graph data with an adjacency list representation,
GBase~\cite{kang11} uses a sparse matrix format. The matrix representation
allows GBase to use compression schemes to store homogenous regions of graphs,
significantly reducing the storage overhead for large graphs. On top of this
storage layout, GBase provides a parallel indexing mechanism that accelerate
queries. While the high-level motivations of GBase (i.e., improving query
response time for graph database queries) are similar to our work, they are
largely focused on the storage overhead. In contrast, we focus on reducing the
query I/O cost.

DeltaGraph~\cite{khurana13}, like our work, includes a temporal component to the
layout design, to efficiently support queries over historical graph
data. DeltaGraph differs from our work in that they are targeting distributed
graph databases, that partition data across a set of machines. Consequently,
they propose a quite different cost model. Moreover, our railway design lays the
foundation for an adaptive disk layout mechanism, which can change over
time. Since the DeltaGraph mechanism is static, we expect that the two designs
are complementary.

Finally, there is prior work on temporal RDF databases, which aims to improve
the response time of SPARQL queries. Notably, Bornea et al.~\cite{bornea13}
describe a way of mapping an RDF store to a relational database, in order to
leverage the overwhelming amount of work on relational database query
optimization. Their work is similar to ours in that they use a ILP formulation
of a constraint problem in order to optimally determine data placement.


\section{Conclusion}\label{sec:conclusion}
\noindent
Many of today's most popular applications rely on data analytics performed on
Interaction graphs. The ability to efficiently support historical analysis over
interaction graphs require effective solutions for the problem of data layout on
disk. In this paper, we have presented a novel disk layout design for graphs
called the railway layout. The design is analogous to hybrid column and row
stores in relational databases. Our simulations show that the railway layout
significantly reduces query I/O cost for randomized workloads. We have
identified the key challenge for systems to implement the railway layout, which
is how to partition blocks into sub-blocks. To solve that problem, we first
presented optimal solutions for overlapping and non-overlapping partitioning
using an ILP formulation. To improve the scalability of the partitioner, and
enable future work in online adaptation of the disk layout, we have also
presented heuristic greedy algorithms that find results close to the optimal
solutions, but exhibit faster running times on large graph schemas and
workloads. To compare the four partitioning algorithms, we have presented a
number of experiments that evaluate the effectiveness and tradeoffs of the
various approaches. Overall, the railway layout design appreciably improves
the performance of data analytics on interaction graphs, and lays the ground
work for future systems design research.

\bibliographystyle{abbrv}
\bibliography{paper}  
\end{document}